\documentclass[a4paper,fleqn,usenatbib]{mnras}

\usepackage{newtxtext,newtxmath}

\usepackage[T1]{fontenc}
\usepackage{ae,aecompl}

\usepackage[dvipdfmx]{graphicx}	
\usepackage{amsmath}	
\usepackage{amssymb}	
\usepackage{natbib}
\usepackage{soul}
\usepackage{lscape}

\newcommand{\spl}{\delta\nu_{\star}}
\newcommand{\is}{i_{\star}}
\newcommand{\cosis}{\cos{i_{\star}}}
\newcommand{\resol}{{\delta}f}
\newcommand{\dnu}{\Delta\nu}
\newcommand{\dnusun}{\Delta\nu_{\odot}}
\newcommand{\numax}{\nu_{\rm max}}
\newcommand{\numaxsun}{\nu_{\rm max, \odot}}

\newcommand{\Teffsun}{T_{\rm eff, \odot}}
\newcommand{\io}{i_{\rm orb}}
\newcommand{\lam}{\lambda}
\newcommand{\vsini}{v\sin{i_{\star}}}
\newcommand{\RM}{Rossiter-McLaughlin\:}

\title[Stellar inclination from asteroseismology]{Reliability of stellar
inclination estimated from asteroseismology: analytical criteria, mock
simulations and Kepler data analysis}

\author[S. Kamiaka et al.]{
Shoya Kamiaka,$^{1}$\thanks{E-mail: kamiaka@utap.phys.s.u-tokyo.ac.jp}
Othman Benomar,$^{2}$
and Yasushi Suto$^{1,3}$
\\
$^{1}$Department of Physics, The University of Tokyo, Tokyo 113-0033, Japan\\
$^{2}$Center for Space Science, NYUAD Institute, New York University Abu Dhabi, PO Box 129188, Abu Dhabi, UAE\\
$^{3}$Research Center for the Early Universe, The University of Tokyo, Tokyo 113-0033, Japan
}

\date{Accepted XXX. Received YYY; in original form ZZZ}

\pubyear{2017}

\begin{document}
\label{firstpage}
\pagerange{\pageref{firstpage}--\pageref{lastpage}}
\maketitle

\begin{abstract}
Advances in asteroseismology of solar-like stars, now provide a unique
method to estimate the stellar inclination $\is$.  This enables to
evaluate the spin-orbit angle of transiting planetary systems, in a
complementary fashion to the \RM effect, a well-established method to
estimate the {\it projected} spin-orbit angle $\lambda$.  Although the
asteroseismic method has been broadly applied to the {\it Kepler}
data, its reliability has yet to be assessed intensively.  In this
work, we evaluate the accuracy of $\is$ from asteroseismology of
solar-like stars using 3000 simulated power spectra. We find
that the low signal-to-noise ratio of the power spectra induces a
systematic under-estimate (over-estimate) bias for stars with high
(low) inclinations.  We derive analytical criteria for
the reliable asteroseismic estimate, which indicates that reliable
measurements are possible in the range of $20^\circ \lesssim \is
\lesssim 80^\circ$ only for stars with high signal-to-noise ratio.
We also analyse and measure the stellar inclination of 94 {\it Kepler}
main-sequence solar-like stars, among which 33 are planetary
hosts. According to our reliability criteria, a third of them (9 with
planets, 22 without) have accurate stellar inclination.  Comparison of
our asteroseismic estimate of $\vsini$ against spectroscopic
measurements indicates that the latter suffers from a large
uncertainty possibly due to the modeling of macro-turbulence,
especially for stars with projected rotation speed $v\sin \is \lesssim
5$km/s. This reinforces earlier claims, and the stellar
inclination estimated from the combination of measurements from spectroscopy and photometric variation for
slowly rotating stars needs to be interpreted with caution.
\end{abstract}




\section{Introduction
\label{sec:intro}}

Asteroseismology, the science that studies pulsation of stars, is a
powerful tool to explore the stellar internal structure. Since
it requires to observe bright stars over long period of time, however, its
applicability has been rather limited, and the Sun has been the major
target for decades.

The situation has drastically changed recently, thanks to the space
missions, {\it CoRoT} \citep{Baglin2006b, Baglin2006a} and {\it Kepler}
\citep{Borucki2010}, which enabled dedicated photometric long-term
monitoring of hundreds of thousands of stars.  Asteroseismology is
known to significantly improve the precision of fundamental parameters
of stars, down to a few percents in radius and $\simeq 5$ percents in
mass, respectively. Also asteroseismology helps improve the age
estimate of stars down to $\simeq 15$ percents when combined with
stellar modeling \citep{Silva-Aguirre:2017aa}.

This is why asteroseismology is becoming an important method to study
stellar populations.  For example, asteroseismology allows to provide
a new insight on the formation and evolution history of the milky way
\citep{Casagrande2016,Miglio2016,Miglio2017}.  Asteroseismology can
also play an essential role in current and future surveys of
exoplanets \citep{Davies:2016aa,Plato2016,VanEylen2017}.  Indeed, the
most basic parameters such as the mass and radius of exoplanets, are
usually estimated on the basis of those from their host stars.
Thus
the precise determinations of the fundamental parameters such as mass and radius of host stars are
essential.

Another important quantity that asteroseismology can also deliver
uniquely is the inclination angle of stellar spin axis with
respect to our line-of-sight, $\is$.  Indeed the correlation between the
stellar spin vector $\vec{s}$ and the angular momentum vector of
planetary orbit $\vec{l}$ is widely recognized as an important clue to
the initial condition and migration history of exoplanetary systems
\citep[e.g.,][and references therein]{Winn:2015aa}.

We refer to the angle $\psi$ between the stellar spin and 
planetary orbital momenta as the spin-orbit angle.
Observationally $\cos{\psi}$ can be decomposed into the components
perpendicular and parallel to the observer's line-of-sight and expressed
in terms of three angles as
\begin{eqnarray}
\label{eq:cospsi}
\cos{\psi} = \sin{\is}\sin{\io}\cos{\lam} + \cosis\cos{\io},
\end{eqnarray}
where $\io$ is the inclination angle of the planetary orbit, and $\lam$
denotes the {\it projected} spin-orbit angle introduced by
\citet{Ohta:2005aa}.

High-resolution spectroscopic observations of transiting
planetary systems enable to measure $\lam$
\citep{Holt:1893aa,Rossiter:1924aa,McLaughlin:1924aa,Queloz:2000aa}.
Since $\io \simeq \pi/2$ for transiting planets, equation
(\ref{eq:cospsi}) reduces to
\begin{eqnarray}
\label{eq:cospsi2}
\cos{\psi} \simeq \sin{\is}\cos{\lam}.
\end{eqnarray}
Clearly the measurement of $\is$ is essential to recover the true
spin-orbit angle $\psi$, in addition to the projected angle
$\lambda$. This is why asteroseismology is important
\citep{Toutain1993, Gizon:2003aa}.  Indeed $\is$ for planetary host
stars has been measured by several authors, including
\citet{Huber:2013aa}, \citet{Chaplin:2013ab},
\citet{Benomar:2014aa}, and \citet{Campante:2016aa}.  For example,
\citet{Benomar:2014aa} found that $\is \simeq 30^{\circ}$ and $\psi
\simeq 120^{\circ}$ for HAT-P-7, and $\is \simeq 65^{\circ}$ and
$\psi \simeq 30^{\circ}$ for Kepler-25.

The number of transiting exoplanetary systems with measured $\psi$
using asteroseismology is hampered by the required signal-to-noise
ratio (SNR) and the relatively high stellar rotation rate.  In
reality, a majority of transiting exoplanets have been searched around
F, G, and K type stars in their main-sequence phase.  For such
main-sequence stars, $\is$ is difficult to measure.  Their oscillation
modes exhibit low amplitudes and suffer from severe blending among
modes \citep[see e.g.,][]{Appourchaux2008}.  Therefore the systematic
verification of the reliability of $\is$ derived from asteroseismology
is of fundamental importance.  As shown by \citet{Gizon:2003aa} and
\citet{Ballot:2006aa,Ballot:2008aa} using a limited number of
simulations, the inferred value of $\is$ is not so accurate if
modes are of insufficient SNR or not properly identified.  This
is why we attempt here to perform systematic mock simulations to
examine the reliability of $\is$ determination from asteroseismology.
In the present paper, we focus on main-sequence stars, but our
method could be applied to evolved stars as well. This is
because the fundamental physical reason allowing us to infer the
inclination are the same
\citep{Beck2012Nature,Benomar2013a,Mosser2017}.
In the future, it is
worthwhile to extend our methodology for those evolved
stars, as discussed in more details in section \ref{subsec:implications}.
  
Section \ref{sec:model} briefly reviews the basic method of $\is$
determination proposed by \citet{Gizon:2003aa}, and then presents our
approximations in generating mock power spectra.  We consider several
conditions for possible degeneracies in line profile fitting in
section \ref{sec:bias}, and then compare those analytic conditions
against the simulation results in section \ref{sec:simulation}.  In
section \ref{sec:Kepler}, we perform asteroseismic analysis for {\it
Kepler} stars with and without known planetary companions, and
examine their reliability discussed in the previous
section. Finally, section \ref{sec:discussion} discusses the
implications of our findings and section \ref{sec:summary} is
devoted to summary and implications of the present paper.

\section{Determination of stellar inclinations from asteroseismology}
\label{sec:model}

In this section, we briefly describe the basic methodology of estimating
$\is$ from asteroseismology following \citet{Gizon:2003aa}.  In
addition, we summarize several approximations that are commonly
adopted in the real data analysis, and therefore in our mock simulations
below as well.

\subsection{Basic model
\label{subsec:basic_model}}

{\it Kepler} performed photometric monitoring of $\sim 1.5 \times
10^{5}$ stars over 4 years.  For application of asteroseismology, the
{\it Kepler} short cadence data with its one minute exposure is
essential because it allows us to explore the frequency regime of
pulsations for the solar-like stars in the main-sequence phase.  The
time-dependent light curve of each star is transformed into the
corresponding power spectra $P(\nu)$, in which stochastically excited
pulsation modes can be easily identified.  In the frequency domain,
each pulsation mode is characterized by a set of three indices
$(n,l,m)$, corresponding to the spherical harmonics
$Y_{l,m}(\theta,\varphi)$ defined at each radial eigen-mode $n$.  The
indices $n$, $l$, and $m$ are referred to as radial order, angular
degree, and azimuthal order, respectively.  Each oscillation mode is
approximated by a Lorentzian line profile, and is characterized by its
height $H(n,l,m,\is)$, width $\Gamma(n,l,m)$, and central frequency
$\nu(n,l,m)$.  Note that the dependence on $\is$ is entirely imprinted
in its height $H(n,l,m,\is)$ as we describe later.

Thus the entire power spectra $P(\nu)$ are well approximated by a
summation of different oscillation modes:
\begin{eqnarray}
\label{eq:model}
P(\nu) = 
\sum_{n=n_{\rm min}}^{n_{\rm max}}
\sum_{l=0}^{l_{\rm max}}
\sum_{m=-l}^{+l}
\frac{H(n,l,m,\is)}{1+4[\nu-\nu(n,l,m)]^{2}/\Gamma^{2}(n,l,m)} + N(\nu),
\end{eqnarray}
where $N(\nu)$ is a background model.  The background is essentially
due to the convection motion at the stellar surface.  As discussed by
e.g., \citet{Harvey:1985aa,Appourchaux:2009aa,Karoff2013a}, the
background model can be described as a sum of several semi-Lorentzian
and of a white noise background.  Each semi-Lorentzian relates to the
flow motion at different spatial and temporal scales.  The white noise
corresponds to the noise limit, mostly due to the photon shot noise.
Because the pulsation in main-sequence stars occurs at high
frequency (e.g., in the Sun, the so-called 5 min oscillations), it
suffices to consider only two semi-Lorentzians for the background
model:
\begin{eqnarray}
\label{eq:noise-model}
N(\nu) = \frac{A_{1}}{1+(\tau_{1}\nu)^{p_{1}}} 
+ \frac{A_{2}}{1+(\tau_{2}\nu)^{p_{2}}} + N_{0}.
\end{eqnarray}
Here $N_0$ is the white noise and $A$, $\tau$, and $p$ correspond to the
height, characteristic time scale, and slope of semi-Lorentzian,
respectively.

In the rotating frame of each star, the eigen-mode frequency is
independent of $m$, and given by $\nu(n,l)$.  Thus the central frequency
in the observer's frame becomes
\begin{eqnarray}
\label{eq:frequency}
\nu(n,l,m) = \nu(n,l) + m\spl
\simeq \left(n+\frac{l}{2}+ \varepsilon_{n,l}\right)\dnu + m\spl ,
\end{eqnarray}
where $\dnu$ is referred to as a large separation (a frequency spacing
between consecutive radial modes), $\varepsilon_{n,l}$ is a small
correction of order unity \citep[e.g.,][]{Tassoul1980, Tassoul1990,
Mosser2013}, and $\spl$ is approximately the inverse of the
stellar rotational period and called the stellar rotational splitting
\citep[see e.g.,][]{Appourchaux2008}.  Thus the degeneracy among $m$
can be broken due to the stellar rotation.

The height
$H(n,l,m,\is)$ of the mode for the observer is known to be given by
\begin{eqnarray}
H(n,l,m,\is) = \mathcal{E}(l,m,\is) H(n,l),
\end{eqnarray}
where
\begin{eqnarray}
\label{eq:explicit}
\mathcal{E}(l,m,\is) = \frac{(l-|m|)!}{(l+|m|)!}\left[P_{l}^{|m|}(\cosis)\right]^{2},
\end{eqnarray}
and $P_{l}^{|m|}$ is the associated Legendre polynomials with degree $l$
and order $m$ \citep[see][]{Toutain1993, Gizon:2003aa}. For instance,
\begin{eqnarray}
\label{eq:Legendre_explicit}
\mathcal{E}(1,0,\is) &=& \cos^{2}{\is}, \cr
\mathcal{E}(1,\pm 1,\is) &=& \frac{1}{2} \sin^{2}{\is}, \cr
\mathcal{E}(2,0,\is) &=& \frac{1}{4} \left(3\cos^{2}{\is} -1\right)^2, \\
\mathcal{E}(2,\pm 1,\is) &=& \frac{3}{2} \cos^{2}{\is} \sin^{2}{\is}, \cr
\mathcal{E}(2,\pm 2,\is) &=& \frac{3}{8} \sin^{4}{\is}. \nonumber
\end{eqnarray}
Therefore if each $m$-mode associated to the same degree $l$ is properly
identified in $P(\nu)$, the ratio of their heights can be used to
determine $\is$.

\subsection{Conventional approximations}

An asteroseismic analysis requires to identify the indices
$(n,l,m)$ for each mode from the noisy spectra, and then preferably fit many
lines simultaneously to determine the global parameters $\is$ and
$\spl$.  Therefore one has to adopt several approximations in order to
reduce the number of free parameters as much as possible.  We summarize
conventional assumptions often adopted in asteroseismology.

Since it is known for the Sun that height ratio of non-radial modes
($l\not=0$) and radial ($l = 0$) mode is uniform over the range of
pulsation frequency \citep[][and references therein]{Salabert:2011ab},
the intrinsic height of the oscillation for $l\not=0$ is assumed to be
\begin{eqnarray}
\label{eq:hnr-vl-hn0}
H(n,l) = V^2_{l} H(n,l=0),
\end{eqnarray}
where $V^2_{l}$ is referred as the mode {\it visibility} and independent
of the radial orders $n$. We adopt a slightly different sets of values,
($V^2_1$, $V^2_2$)=(1.449, 0.6589) in simulations and (1.447, 0.5485) in
the real data analysis of section \ref{sec:Kepler}, as described later.

We neglect the $m$-dependence of $\Gamma(n,l,m)$, and the remaining
$l$-dependence is empirically modeled from the set of the fitted values
of $\Gamma(n,l=0)$ as follows.  First we identify the modes $(n,0)$ for
$n_{\rm min} \leq n \leq n_{\rm max}$ from the spectra, and obtain the
corresponding eigen-frequency $\nu(n,0)$ and mode width $\Gamma(n,0)$.  Then we
construct a continuous function $F$ that linearly interpolates those
discrete sets of parameters, {\it i.e.,} $\Gamma(n,0) = F(\nu(n,0))$.
Since earlier analysis of the Sun shows that the point
$(\nu(n,l),\Gamma(n,l))$ stays approximately at the same trajectory of
$(\nu(n,0),\Gamma(n,0))$ \citep{Toutain1992,Garcia:2004aa}, one simply
replaces $\Gamma(n,l\neq0)$ by $F(\nu(n,l\neq0))$ evaluated at the
fitted value of the eigen-frequency $\nu(n,l\neq0)$.

In summary, under the above assumptions, the free parameters
characterizing the entire power spectra include $H(n,l=0)$,
$\Gamma(n,l=0)$, $\nu(n,l)$, the global parameters responsible for the
shape of peaks ($\is$, $\spl$), and background parameters ($A_{1,2}$,
$\tau_{1,2}$, $p_{1,2}$, and $N_{0}$).  Accordingly, the total number
of fitting parameters is $(n_{\rm max}-n_{\rm min}+1)(l_{\rm max}+3) +
9$.

\section{Analytic criteria to distinguish among different azimuthal orders}
\label{sec:bias}

As suggested in the previous section, accurate measurement of $\is$
crucially depends on the ability of identifying the frequencies and
heights of different $m$-modes associated with the same degree $l$.
Ideally, the higher amplitude and the wider separation between
different $m$-modes are required.  More specifically, the former is
represented by the ratio of the height $H(n,l,m,\is)$ and the noise
level, and the latter, by the ratio of the stellar rotation splitting
and the width, $\spl/\Gamma(n,l)$.  This consideration may be translated
into analytic criteria that are necessary to distinguish among different
$m$-modes.

Because $l=0$ modes are insensitive to either rotation ($\spl$) or 
inclination ($\is$), $\is$ can be determined by non-radial modes
($l\neq0$).  For a majority of main-sequence stars whose pulsations are
detected, their visible modes are limited up to $l=2$. Moreover, the
amplitudes of $l=1$ modes are roughly three times larger than those of
$l=2$ modes. Thus $l=1$ modes dominate the ability to determine $\is$ in
practice, and we consider analytic criteria to separate $m=0$ and $m=\pm1$
modes for $l=1$.

\begin{figure*}
\includegraphics[width=2\columnwidth, angle=0]{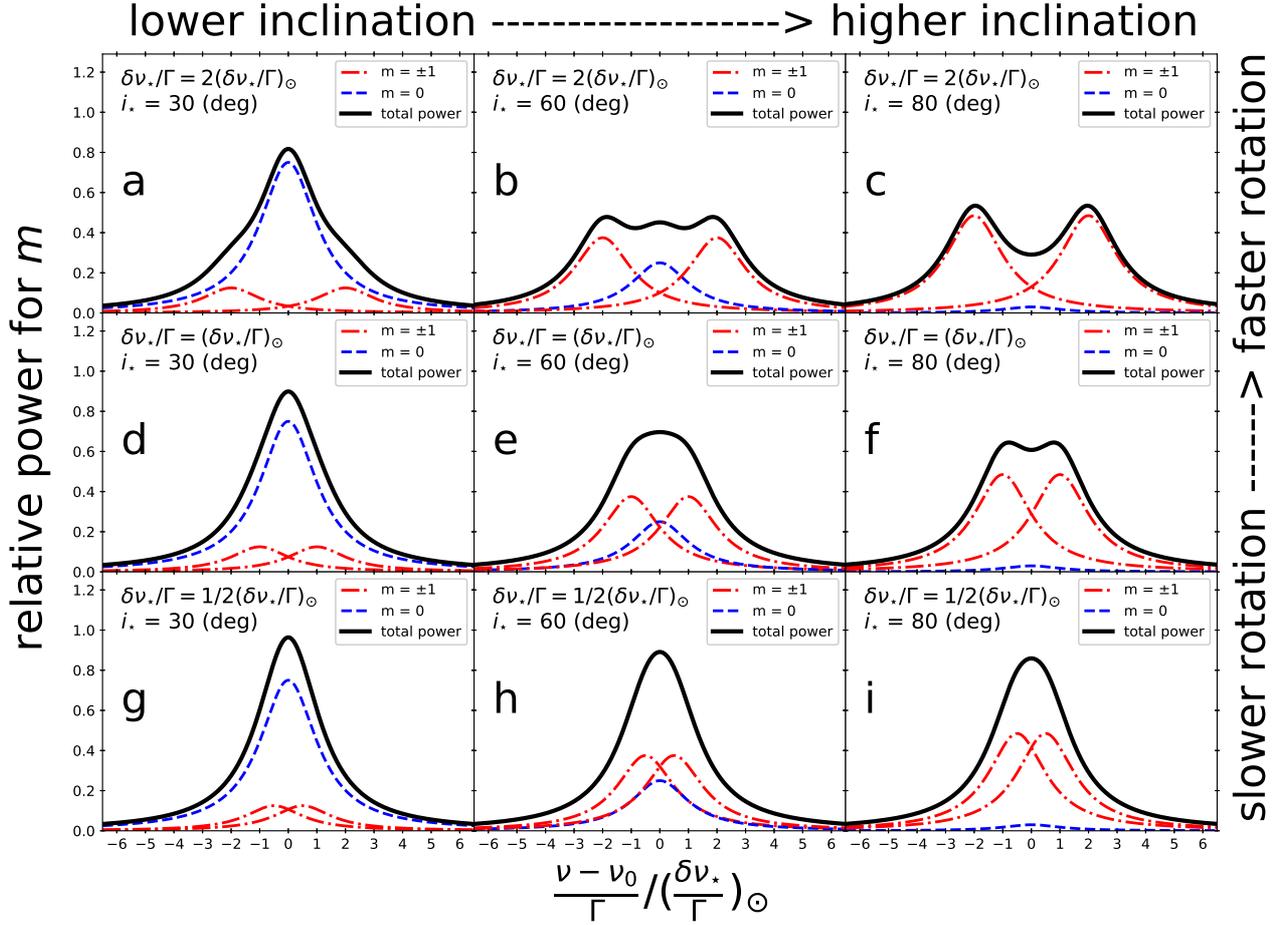}
\caption{
Model power spectra of dipole mode with central frequency
$\nu_{0}$.  $x$ and $y$- axes correspond to the frequency in units
of $(\spl/\Gamma)_{\odot}$ and relative power for different $m$
components in units of the height of $m = 0$ mode for $\is =0$
(deg), respectively.  Blue dashed (red dotted-dashed) line
represents the contribution of $m = 0$ ($m = \pm1$) modes, and black
solid line demonstrates the consequent total power spectra.
Different panel corresponds to the different combination of
$\spl/\Gamma$ and $\is$, ranging from faster (upper) to slower
(lower) rotation and from lower (left) to higher (right)
inclination.
}
\label{fig:mode_appearance_l1}
\end{figure*}

A difficulty to distinguish among different $m$-modes may be understood
from Figure \ref{fig:mode_appearance_l1}, in which model profiles of power
spectra around the central frequency $\nu_0$ for different values of
$\is$ and $\spl/\Gamma$ are plotted; $\is=30^{\circ}$, $60^{\circ}$, and
$80^{\circ}$ from left to right panels, and
$(\spl/\Gamma)/(\spl/\Gamma)_{\odot} = 2$, $1$, and $0.5$ from top to
bottom panels, with $(\spl/\Gamma)_{\odot} \simeq 0.42\mu{\rm
Hz}/0.95\mu{\rm Hz} \simeq 0.44$ being the solar value near the maximum of mode amplitude.  The horizontal
axis corresponds to $(\nu-\nu_0)/\Gamma$ in units of
$(\spl/\Gamma)_{\odot}$.

Hereafter, we note the contribution of $m = 0$ to the power spectra, $P_{l=1,m=0}(\nu)$.
The contribution of $m = \pm1$ modes is noted $P_{l=1,m=\pm1}(\nu)$.
From equations (\ref{eq:model}), (\ref{eq:frequency})
$\sim$ (\ref{eq:Legendre_explicit}),
$P_{l=1,m=0}(\nu)$ and  $P_{l=1,m=\pm1}(\nu)$ are explicitly written as
\begin{align}
\label{eq:l=1m=0}
P_{l=1,m=0}(\nu) &= 
H(n,l=1)\frac{\cos^2\is}{1+4(\nu-\nu_0)^{2}/\Gamma^{2}}, \\
\label{eq:l=1m=pm1}
P_{l=1,m\pm 1}(\nu) &= H(n,l=1)
\frac{\sin^2\is}{2[1+4(\nu-\nu_0 \mp \spl)^{2}/\Gamma^{2})]}.
\end{align}
Figure \ref{fig:mode_appearance_l1} is normalized so that $H(n,l=1)$ is
unity.

The reliability of the estimate of $\is$ and $\spl$ is crucially
determined by how well one can separate the contributions from three
different $m$-modes embedded in the total profile (black solid curve in
Figure \ref{fig:mode_appearance_l1}).  More specifically, their separate
contributions to the total power are computed as
\begin{eqnarray}
\label{eq:powerl=1m=0}
\int_{0}^{\infty}{\rm d}{\nu}P_{l=1,m=0}(\nu) 
&\simeq& \frac{H(n,l=1)\Gamma(n,l=1)}{2}\cos^2{\is} , \\
\label{eq:powerl=1m=pm1}
\int_{0}^{\infty}{\rm d}{\nu}P_{l=1,m=\pm1}(\nu) 
&\simeq& \frac{H(n,l=1)\Gamma(n,l=1)}{4}\sin^2{\is} .
\end{eqnarray}
These need to be much larger than the {\it resolvable element} of the
power, which is roughly given by the product of the rms noise level
$\sigma_n$ in the observed power spectra and the frequency resolution
$\resol \simeq 1/T_{\rm obs}$ with $T_{\rm obs}$ being the total
observation duration.

The above consideration leads to the following qualitative but analytic
criteria.

\begin{description}
\item[(I)] The identification of $m=0$ mode requires
\begin{eqnarray}
\label{eq:condition-A}
\frac{H(n,l=1)\Gamma(n,l=1)}{2}\cos^2{\is} > \alpha \sigma_n\resol, 
\end{eqnarray}
where we introduce a fudge constant $\alpha$ that will be empirically
determined later through the comparison against mock simulation results.
The condition (\ref{eq:condition-A}) becomes
\begin{eqnarray}
\label{eq:condition-A-2}
\cos^{2}{\is} > \alpha \frac{2}{\rm SNR}\frac{\resol}{\Gamma},
\end{eqnarray}
where we define the {\it signal-to-noise ratio} SNR:
\begin{eqnarray}
{\rm SNR} \equiv \frac{H(n,l=1)}{\sigma_n}.
\end{eqnarray}
Then the inequality (\ref{eq:condition-A-2}) leads to an upper limit 
on the detectable $\is$;
\begin{eqnarray}
\label{eq:condition-A-3}
\is < \cos^{-1}\sqrt{\alpha \frac{2}{\rm SNR}\frac{\resol}{\Gamma}}.
\end{eqnarray}
In other words, one cannot reliably estimate the true value of $\is$ if
it is larger than the threshold value in the right-hand-side of the
above inequality. For instance, a reliable estimate of $\is=90^{\circ}$
is very demanding and requires an ideal observation with either
SNR$=\infty$ or $\resol=0$.

\item[(II)] Similarly the identification of $m=\pm1$ mode requires
\begin{eqnarray}
\label{eq:condition-B}
\frac{H(n,l=1)\Gamma(n,l=1)}{4}\sin^2{\is}  > \beta \sigma_n\resol,
\end{eqnarray}
with $\beta$ being another fudge factor to be estimated later.  In this
case, one obtains a lower limit on the measurable $\is$ as
\begin{eqnarray}
\label{eq:condition-B-2}
\is > \sin^{-1}\sqrt{\beta \frac{4}{\rm SNR}\frac{\resol}{\Gamma}}.
\end{eqnarray}
Again this condition implies that either SNR$=\infty$ or $\resol=0$ is
needed for $\is$ to be measurable down to $0^{\circ}$.
\end{description}

The above two criteria are independent of the rotational splitting $\spl$.
Indeed if we consider conditions to distinguish among the peak height of
different $m$-modes, instead of their total area (i.e., power), we obtain
criteria dependent on $\spl/\Gamma$. In this case, however, the
dependence on the frequency resolution $\resol$ is neglected.  In
principle, we could combine those conditions to improve our analytic
criteria, but the results become a bit complicated.  Therefore we decide
to consider the condition on $\spl$ separately as follows.

\begin{description}
\item[(III)] The requirement that the peaks of $m=\pm1$ modes need to be
distinguished from the sum of the $m=\pm1$ modes at $\nu = \nu_{0}$ is
written as
\begin{eqnarray}
\label{eq:condition-C}
P_{l=1,m\pm1}(\nu_0 \pm \spl) >
[P_{l=1,m=+1}(\nu_0) + P_{l=1,m=-1}(\nu_0)],
\end{eqnarray}
which simply reduces to
\begin{eqnarray}
\label{eq:condition-C-2}
\frac{\spl}{\Gamma} > 0.5.
\end{eqnarray}
This is insensitive to $\is$, and a very generic requirement on
$\spl/\Gamma$ to reliably estimate $\is$.  We note that interestingly
the solar value $\simeq 0.44$ is very close to this threshold just by
chance.
\end{description}

\section{Mock simulation to extract stellar inclinations from oscillation power spectra}
\label{sec:simulation}

\begin{figure*}
\includegraphics[width=\columnwidth,angle=0]{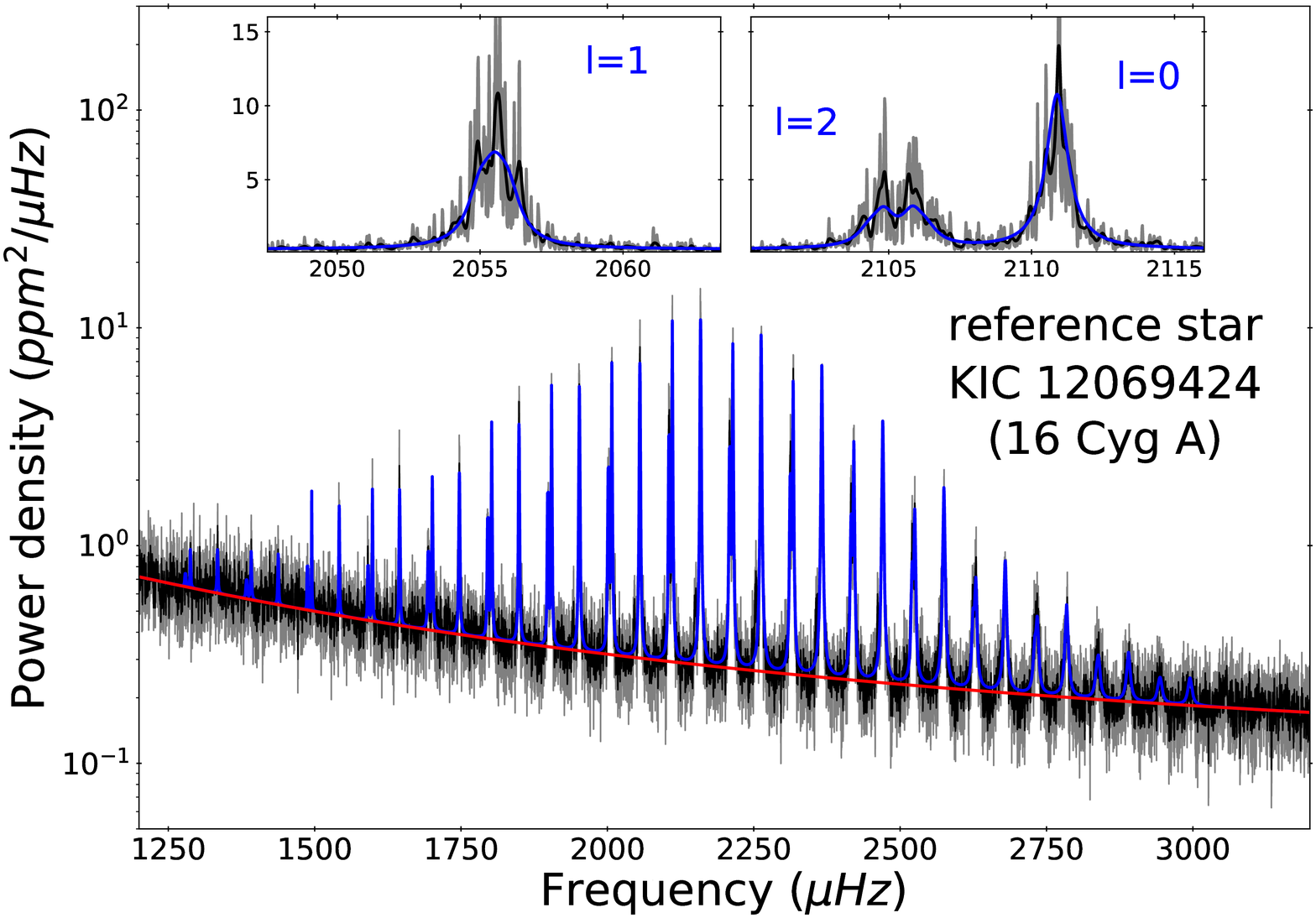}
\includegraphics[width=0.92\columnwidth,angle=0]{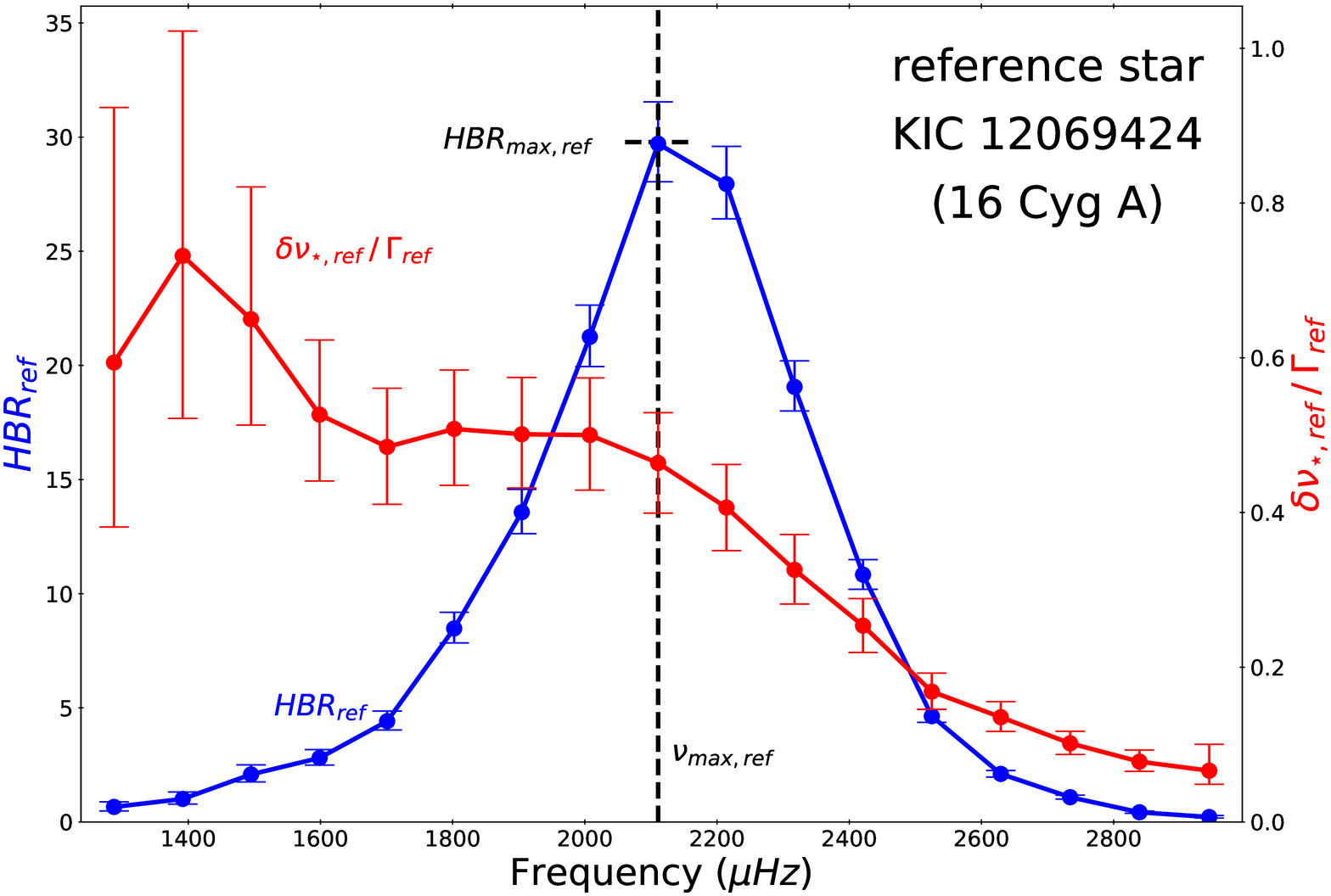}
\caption{
Observed properties of our reference star KIC 12069424 (16
Cyg A).  {\it Left.} Seismic power spectrum smoothed using a
Gaussian filter of width $0.05\,\Delta\nu \simeq 5.2\mu$Hz (gray)
and $0.15\,\Delta\nu \simeq15.5\mu$Hz (black). Superimposed is the
background level (red) and the fitted oscillation frequency
(blue). The inset shows the modes of degree {\it l}=0,1,2 of
highest amplitude.  {\it Right.} Measured
height-to-background (HBR; blue) and splitting-to-width ratio
($\spl/\Gamma$; red) of the radial modes, as a function of the
oscillation frequency $\nu(n,l=0)$. These reference profiles are
scaled and used in simulated power spectra discussed in Section
\ref{subsec:sim_setups}.
}
\label{fig:16CygA_profile}
\end{figure*}

Measurement of the stellar inclination from asteroseismology is based on
several complicated procedures and their validity can be examined
quantitatively only through the analysis of simulated power spectra. We
carry out intensive analyses of systematic mock spectra. We first
describe how to generate simulated power spectra, and then present the
results against our analytic criteria discussed in the previous
section.

\subsection{Generating mock power spectra scaled
from a reference star KIC 12069424
\label{subsec:sim_setups}}

As we demonstrated in the previous section, the precision and accuracy
of the estimate of the stellar inclination depend sensitively on the
splitting-to-width ratio $\spl/\Gamma$ and on the
signal-to-noise ratio SNR;
in the present simulation, we have not
implemented realistic noises except for the background noise level of
$N_0$ in equation (\ref{eq:noise-model}). Thus the signal-to-noise
ratio SNR $\equiv H(n,l=1)/\sigma_n$ defined in section \ref{sec:bias} is not easy
to properly assign. Instead, we use height-to-background ratio HBR $\equiv H(n,l)/N_{0}$ as a proxy for SNR
throughout the following analysis.  In practice, the difference
between HBR and SNR is expected to be incorporated by renormalizing
the values of $\alpha$ and $\beta$ in inequalities
(\ref{eq:condition-A-3}) and (\ref{eq:condition-B-2}).

We take the HBR and $\spl/\Gamma$ as our primary
variables in simulated power spectra, and generate realistic mock
spectra covering a wide range of their values. In practice, we first
choose KIC 12069424 (16 Cyg A) as our reference star, which is one of
the brightest stars monitored by {\it Kepler}. It has one of the
highest HBR among the observed main-sequence stars, and therefore is
one of the most studied star in the {\it Kepler} field
\citep{Metcalfe2012,Metcalfe:2014aa,Davies:2015aa}. The observed
power spectrum for KIC 12069424 is shown in the left panel of Figure
\ref{fig:16CygA_profile}.

We extract the mode parameters (frequency, height, width, noise
background parameters, rotational splitting and inclination) of the
reference star by fitting a sum of Lorentzians and a noise profile as
equation (\ref{eq:model}) in section \ref{subsec:basic_model}.  We use
an MCMC sampling method based on an updated version to C++ of the
algorithm from \citet{Benomar:2009ab}. We fit a total of 17 radial
orders with associated degrees $l=0$, $1$, and $2$.  While $l=3$ degree is
identifiable for this reference star, it is not the case for a
majority of stars. Therefore we do not incorporate the modes of $l
\geq 3$ for the reference star and for the simulated spectra.  We
verified that frequencies, rotational splitting and stellar inclination are all
consistent with the result derived by \citet{Davies:2015aa} within
$2\sigma$.  The right panel of Figure \ref{fig:16CygA_profile} shows
the measured profile of HBR (HBR$_{\mathrm{ref}}$) and the
splitting-to-width ratio
($\delta\nu_{\star,\mathrm{ref}}/\Gamma_{\mathrm{ref}}$) of KIC 12069424 as
a function of the radial mode frequency.  These profiles are fairly
representative of other solar-like stars \citep[see e.g.,][]{Appourchaux2012}.

As shown in the right panel of Figure \ref{fig:16CygA_profile}, the
mode height-to-background ratio $\mathrm{HBR}(n,l=0)$ as a function of
the radial mode frequency has a peak at $\nu_{\rm max, ref}$, and the
corresponding peak value of HBR is defined as $\mathrm{HBR}_{\rm max, 
ref}$.  We scale those reference parameters to generate mock spectra
covering a range of HBR and $\spl/\Gamma$ as described below.
Hereafter, the subscript $\star$ indicates the variables of simulated stars scaled from a reference star.

The mode heights $H_\star(n,l=0)$ of a simulated star are specified by
its maximum value of HBR at $\nu_{\rm max,\star}$; $\mathrm{HBR}_{\rm
max,\star}$, and are scaled as
\begin{eqnarray}
\label{eq:hnr-star-n0}
\mathrm{H_{\star}}(n,l=0) &=& \mathrm{HBR_{\star}}(n,l=0)\,N_0 \nonumber \\
&=& \mathrm{\frac{HBR_{max,\star}}{HBR_{max,ref}} HBR_{ref}}(n,l=0)\,N_0.
\end{eqnarray}
In reality, the noise background of the spectrum of actual
main-sequence solar-like stars weakly depends on frequency (see
Section \ref{subsec:basic_model}). It typically decreases by a
factor of a few between modes with the lowest measurable frequency
and those with the highest frequency. The left panel of Figure
\ref{fig:16CygA_profile} illustrates this for KIC 12069424 (red
solid line). Here, for simplicity we neglect the frequency
dependence in equation (\ref{eq:hnr-star-n0}).  We consider only the
white noise $N_0$, and adopt the value of the reference star $N_{0,
{\rm ref}}=0.0857 {\rm ppm}^2/{\mu {\rm Hz}}$. The height
$\mathrm{H_{\star}}(n,l)$ for $l=1$ and $l=2$ are scaled using the
visibilities of the reference star, $V^2_{1, {\rm ref}}=1.449$
and $V^2_{2, {\rm ref}}=0.659$, from equations
(\ref{eq:hnr-vl-hn0}) and (\ref{eq:hnr-star-n0}).

The other primary parameter that controls the reliability of the
estimate of $\is$ is the splitting-to-width ratio
$\spl/\Gamma(n,l)$. It measures the influence of the overlap between
split components (see section \ref{sec:bias}) on the inclination. In the
current simulation, we fix the width of the mode $\Gamma(n,l)$ to the
reference value $\Gamma_{\mathrm{ref}}(n,l)$. On the other hand, we modify the
rotational splitting $\spl$. Then,
\begin{eqnarray}
\spl = \gamma_{\star} \Gamma_{\mathrm{max, ref}},
\end{eqnarray}
where $\Gamma_{\mathrm{max, ref}} =1.08\mu \mathrm{Hz}$ is the
width of the mode that corresponds to HBR$_{\mathrm{max,ref}}$. Here, $\gamma_{\star}$ is the splitting-to-width ratio at $\mathrm{HBR_{max,\star}}$.

Obviously the observation duration $T_{\mathrm{obs}}$ is another
important factor that defines the number of independent data points
sampling a mode profile.  The longer $T_{\mathrm{obs}}$ improves the
frequency resolution $\resol \propto 1/T_{\mathrm{obs}}$.  It also improves the description of the mode profile, which in turns enhances the accuracy on the stellar inclination.
To assess this, we consider
$T_{\mathrm{obs}}=1$ and 4 years, corresponding to the minimal and
maximal observation duration of the LEGACY \footnote{The LEGACY
sample corresponds to the ensemble of {\it Kepler} main-sequence stars that show Sun-like pulsations and that were observed continuously for at least a year.}
{\it Kepler} sample \citep{Lund:2017aa}.

\begin{table}
\label{tab:grid}
\caption{Range of parameter values in simulated spectra 
with $T_{\mathrm{obs}}=1$ year and $T_{\mathrm{obs}}=4$ years.}
\centering
\begin{tabular}{ccc}
parameter & range & grid interval \\
\hline
$\mathrm{HBR_{max,\star}}$ & [0,30] & 1 \\
$\mathrm{\gamma_{\star}}=\delta\nu_\star/\Gamma_{\mathrm{max, ref}}$ 
& [0.1, 1.0] & 0.1 \\
$\is$(deg) & [0, 90] & 10 \\
\hline
\end{tabular}
\end{table}

A grid with a total of $3000$ artificial spectra is generated each for
$T_{\mathrm{obs}}=1$ and 4 years.  Table \ref{tab:grid} summarizes the
ranges of the three control parameters for the simulated spectra;
HBR$_{\rm max,\star}$, $\gamma_{\star}$ and the inclination angle
$\is$.

\begin{figure*}
\includegraphics[width=\columnwidth]{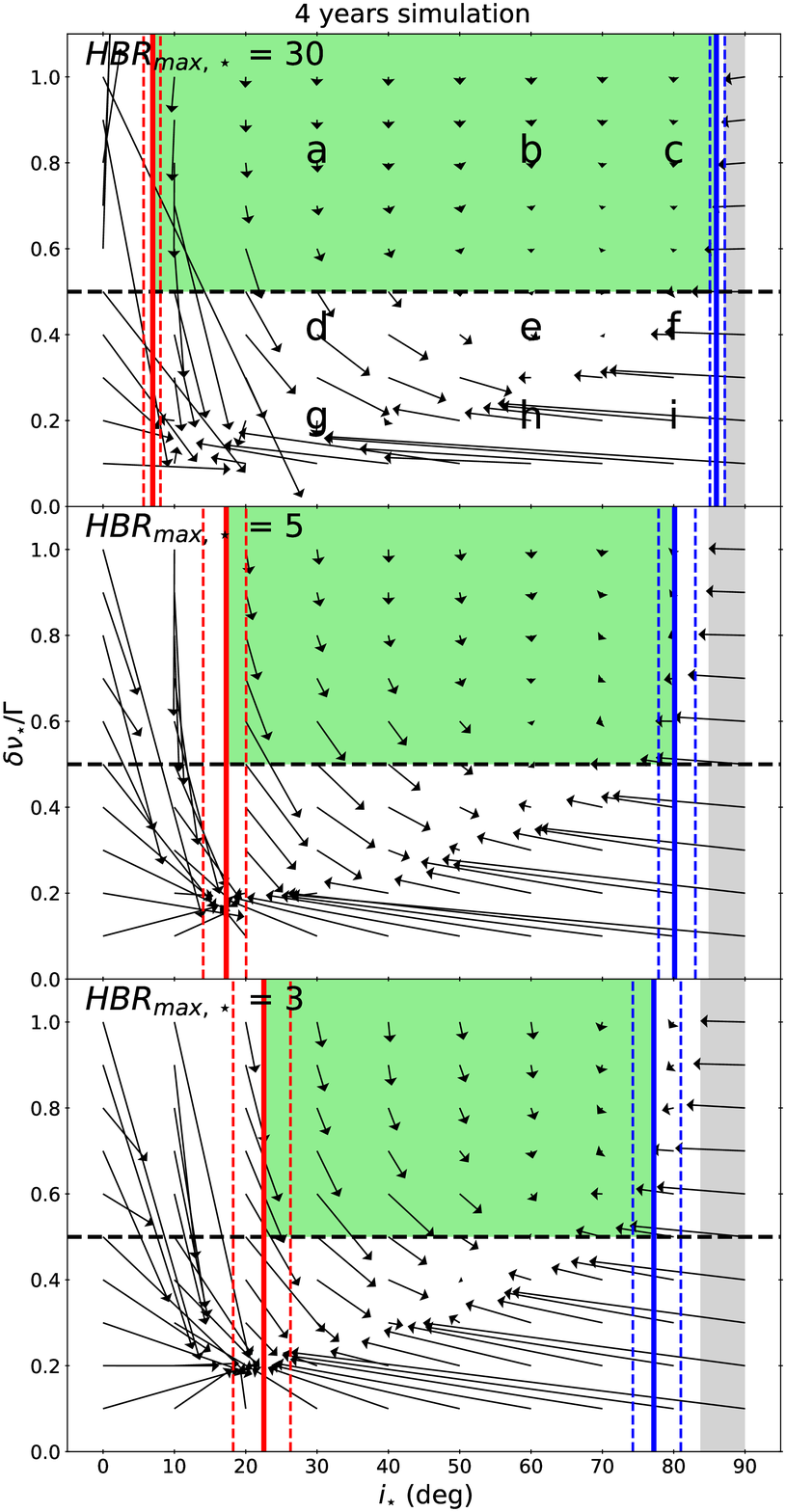}
\includegraphics[width=\columnwidth]{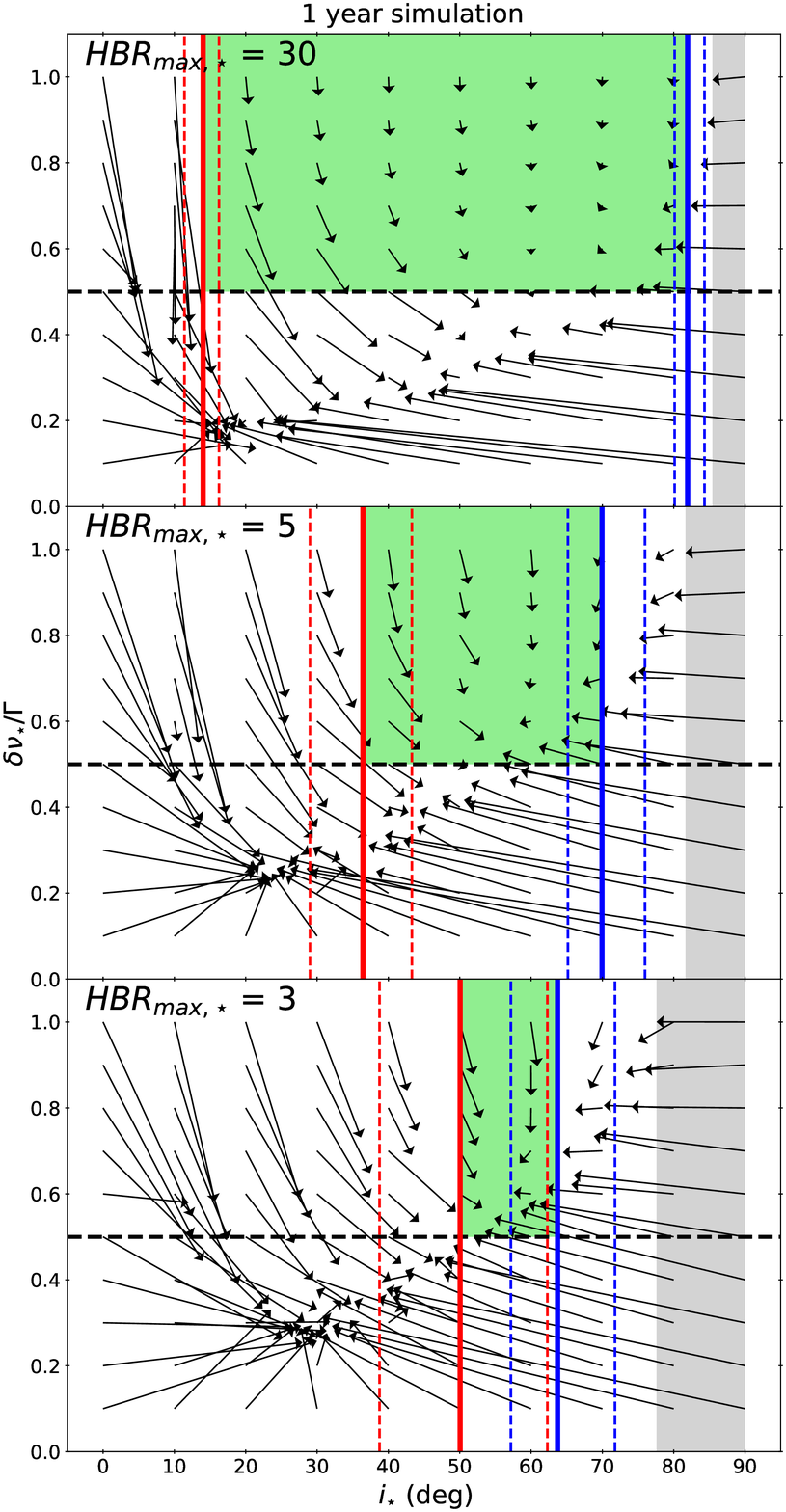}
\caption{
Comparison of asteroseismically derived values and true inputs
of $\is$ and $\spl/\Gamma$ for simulated spectra. Arrows in each panel
start from the true values and end at the estimated values.  Left and
right panels show the results for $T_{\rm obs}=4$ years and 1 year,
respectively.  Top, middle, and bottom panels correspond to HBR$_{\rm
max, \star}$ = 30, 5, and 3.  Nine alphabets in the top-left panel
indicate the set of ($\is$, $\spl/\Gamma$) assumed in the the
corresponding panel of Figure \ref{fig:mode_appearance_l1}.  Blue lines, red lines and horizontal
black dashed lines correspond to analytic criteria (I), (II), and (III)
discussed in Section \ref{sec:bias}, respectively.
}
\label{fig:bias}
\end{figure*}

\subsection{Results of mock spectra analysis
\label{subsec:sim_results}}

\begin{figure*}
\includegraphics[width=1.5\columnwidth]{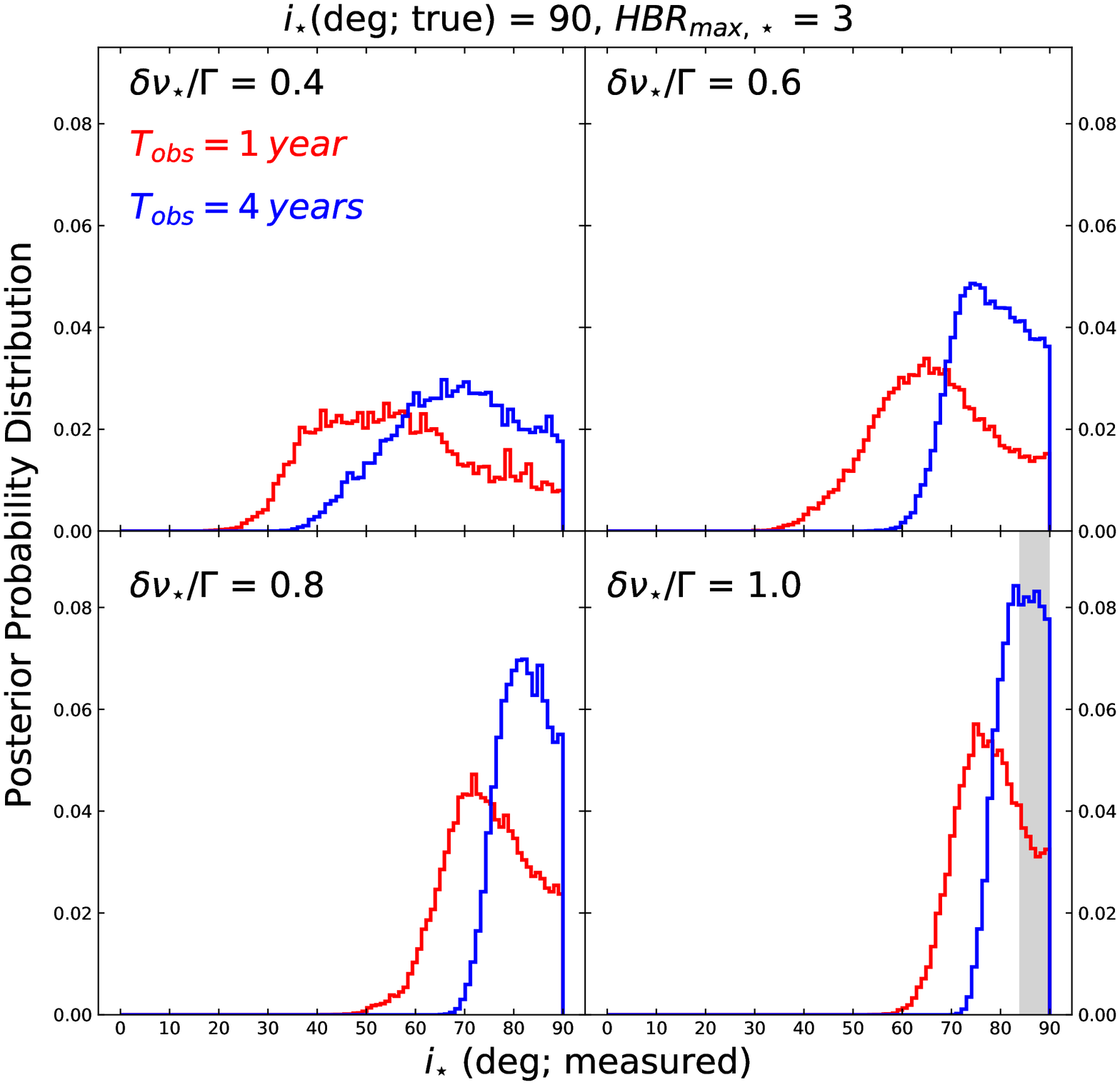} 
\caption{
Posterior probability distribution (PPD) of measured $\is$
for simulated spectra with input $\is=90^\circ$ with different input
$\spl/\Gamma$. HBR$_{\rm max, \star}=3$ is assumed.
These are shown as histograms, with the number of bins defined according to Scott's normal reference rule \citep{Scott:1992aa}.
Red and blue histograms show the results for $T_{\rm
obs}=1$ year and 4 years, respectively.  Gray area in the lower-right
panel indicates the region from the median of the 4 year PPD ($\is =
83.8^{\circ}$) to the input value ($\is = 90^{\circ}$).
}
\label{fig:1yr_to_4yr}
\end{figure*}

Figure \ref{fig:bias} plots the result of mock spectra analysis on
$\is$ - $\spl/\Gamma$ plane.  Specifically, it shows the difference
between the true input value and the median of the inferred posterior probability
distribution (PPD) for $\is$ and $\spl/\Gamma$. As we will show
later, the median value does not necessarily represent the best-fit, but
we use it here just for simplicity.  The base of the black arrows
indicates the input value, the tip is the measured median value.  Left
panels are for $T_{\mathrm{obs}}=4$ years, while right panels are for
$T_{\mathrm{obs}}=1$ year.  Top, middle, and bottom panels correspond to
HBR$_{\mathrm{max},\star} = 30$, 5, and 3, respectively.  Note that
HBR$_{\mathrm{max},\star}= 3 \sim 5$ are representative of the maximum
HBR of the modes for {\it Kepler} stars with pulsations.  In practice,
below HBR$_{\mathrm{max}}$ = 3 the noise makes difficult to observe the
individual pulsation modes, so that the seismic analysis is often
limited to the measure of the central frequency at maximum power
$\nu_{\mathrm{max}}$ and of the large separation $\dnu$.  The case with
HBR$_{\mathrm{max},\star}$ = 30 corresponds to the best cases, such as KIC
12069424 (the reference star).

Clearly there exists a coherent pattern of arrow distribution over the
plane, indicating the presence of the systematic bias in the parameter
estimation.  The length of each arrow reflects the amplitude of the
bias.  Labels of ``a'' to ``i'' in the top-left panel of Figure \ref{fig:bias}
indicate the locations of ($\is$, $\spl/\Gamma$) in the corresponding
panels of Figure \ref{fig:mode_appearance_l1}. The comparison of Figure \ref{fig:bias} with
Figure \ref{fig:mode_appearance_l1} helps intuitive understanding of the result.  

To proceed further, we overlay the analytic criteria (I) $\sim$ (III) on
each panel of Figure \ref{fig:bias}.  The three vertical lines in the right
part indicate the criterion (I) with $\alpha=15$ (left dashed), 10
(middle solid) and 5 (right dashed).  Similarly the three vertical
lines in the left part indicate the criterion (II) with $\beta=10$
(left dashed), 15 (middle solid) and 20 (right dashed).  Finally the
horizontal dashed line corresponds to the criterion (III),
$\spl/\Gamma=0.5$. In doing so, we set SNR=HBR$_{\mathrm{max}}$ just for simplicity. 
As we remarked, those criteria are not expected to be
strict, and the adopted values of $\alpha$ and $\beta$ are merely
empirical. Nevertheless the regions bounded by the criteria agree with
those in which the input parameters are reproduced fairly accurately
from the mock simulation.  We also note that the length of the arrows
for $T_{\mathrm{obs}}=4$ years becomes approximately half with respect
to that for 1 year on average.  This indicates that not only the
uncertainty but also the accuracy of the estimate scales as
$1/\sqrt{T_{\mathrm{obs}}}$.

On the basis of the above empirical comparison, we divide the observed
{\it Kepler} stars into two different categories adopting $\alpha = 10$ and
$\beta = 15$ in the next section.

Before finishing this section, it is worth emphasizing the limitation of
using the median value of the derived PPD in estimating the inclination
angle.  For that purpose, we compute the derived PPD of the inclination
angle for a simulated star assigned $\is = 90^{\circ}$ as a true
input.  The resulting PPDs are plotted in Figure \ref{fig:1yr_to_4yr} for
different values of $\spl/\Gamma$.  Red and blue histograms correspond
to 1 year and 4 years simulations, respectively, for HBR$_{\mathrm{max},\star}$ = 3.
While the true value ($90^{\circ}$) can be measured better for higher
$\spl/\Gamma$ and longer observation, the measured value always becomes
less than $90^{\circ}$.  For example, the gray regions in the
bottom-right panel brackets between the measured and true values.
This is because we employ the sampling method for inclination over the
range of [$0^{\circ},90^{\circ}$].  Similarly, the derived inclination
for true inclination of $0^{\circ}$ is always greater than $0^{\circ}$.
Therefore the bias indicated in Figure \ref{fig:bias} may be partly, even
though not entirely, due to the use of the median value of the entire
PPD.

\section{Application to Kepler data}
\label{sec:Kepler}

\begin{figure*}
\includegraphics[width=1.5\columnwidth]{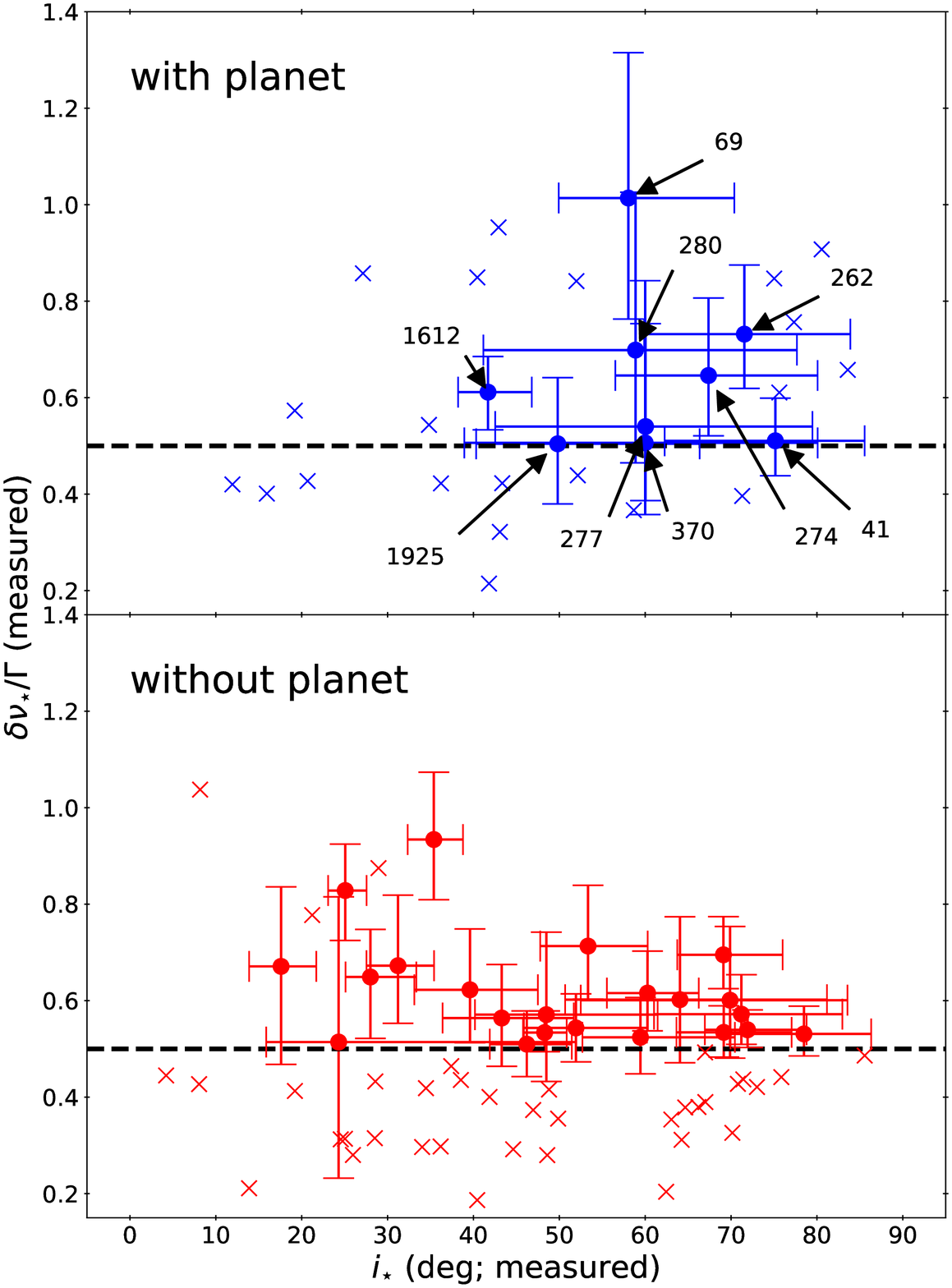}
\caption{
Measured values of $\is$ and $\spl/\Gamma$ for {\it Kepler}
stars with (top) and without (bottom) known planetary companions.
The black solid horizontal line represents
$\spl/\Gamma = 0.5$ (equation \ref{eq:condition-C-2}).  Filled circles
indicate category A stars, while crosses correspond to category B
stars. The numbers labelling the filled circles denote
the KOI IDs for stars with planets.
}
\label{fig:KOI_scatter}
\end{figure*}

\subsection{Target star selection
\label{subsec:Kepler_data}}

We analyse stars monitored by {\it Kepler} during its initial 4 years
mission. In total, we consider 33 stars with transiting planets, and 61
stars without known transiting planets.

The stars without known planets in this work are taken from LEGACY
sample \citep{Lund:2017aa,Silva-Aguirre:2017aa}.  This sample consists
of 66 main-sequence stars observed in short cadence for at least 1
year by {\it Kepler}. Out of the 66 stars, we select 61 stars that do
not have known planets (the remaining 5 stars with planets are also
analysed below).

We re-analyze 25 stars with planets from \citet{Campante:2016aa}, which
include 4 stars (KIC 3632418, 9414417, 9955598, and 10963065) with
planets from the LEGACY sample. In addition, we analyse 8 stars with
planets whose asteroseismic analysis has not yet been published elsewhere,
including one star (KIC 7296438) from LEGACY sample.

The power spectra prepared using the method of \citet{Handberg:2014aa},
are downloaded from Kepler Asteroseismic Science Operations Center
(KASOC) database (http://kasoc.phys.au.dk). There are two different
spectra available, with and without weighting of the photometric flux
with the flux uncertainty.
We use unweighted spectra for 42 targets since
their weighted spectra are not available. Otherwise we use the weighted
spectra for 51 stars. For KIC 11401755,  the latest weighted spectra are
not available, and we decided to use its unweighted spectra.

We adopt $V^2_{1} = 1.447$ and $V^2_{2} =
0.5485$, the mean of visibility of the Sun in green and red VIRGO/SPM
channels from \cite{Salabert:2011ab}. This averaging might give
visibilities representative of the {\it Kepler} visibilities
\citep{Ballot2011}.

\subsection{Asteroseismic inference of Kepler stars
\label{subsec:Kepler_results}}

\begin{table*}
\caption{Results on {\it Kepler} planet-host stars in category A}
\label{tab:output_pass_KOI}
\begin{tabular}{ccccccc}
\hline
KIC & KOI & {\it Kepler} ID & HBR$_{\mathrm {max}}$ & $\spl/\Gamma$ & $\is$(16\%, 50\%, 84\%) & $\spl$(16\%, 50\%, 84\%)\\
& & & & & (deg) & (${\mu}$Hz)\\
\hline
3544595 & 69 & 93 & 2.25 & 1.01 & [49.9,\:58.0,\:70.4] & [0.42,\:0.49,\:0.56] \\
4141376 & 280 & 1655 & 0.69 & 0.70 & [41.2,\:58.9,\:77.7] & [0.80,\:0.98,\:1.37] \\
6521045 & 41 & 100 & 3.13 & 0.51 & [62.3,\:75.2,\:85.6] & [0.43,\:0.46,\:0.51] \\
8077137 & 274 & 128 & 1.45 & 0.65 & [56.5,\:67.4,\:80.1] & [0.84,\:0.93,\:1.05] \\
8494142 & 370 & 145 & 0.80 & 0.51 & [38.9,\:60.0,\:80.1] & [0.89,\:1.09,\:1.72] \\
9955598 & 1925 & 409 & 2.91 & 0.50 & [40.3,\:49.8,\:66.3] & [0.32,\:0.41,\:0.49] \\
10963065 & 1612 & 408 & 7.79 & 0.61 & [38.2,\:41.7,\:46.8] & [0.88,\:0.99,\:1.08] \\
11401755 & 277 & 36 & 1.19 & 0.54 & [42.5,\:60.0,\:79.5] & [0.53,\:0.65,\:0.84] \\
11807274 & 262 & 50 & 1.61 & 0.73 & [59.9,\:71.6,\:83.9] & [1.41,\:1.52,\:1.70] \\
\hline
\end{tabular}\\
\end{table*}

\begin{table*}
\caption{Results on {\it Kepler} planet-host stars in category B}
\label{tab:output_fail_KOI}
\begin{tabular}{ccccccc}
\hline
KIC & KOI & {\it Kepler} ID & HBR$_{\mathrm {max}}$ & $\spl/\Gamma$ & $\is$(16\%, 50\%, 84\%) & $\spl$(16\%, 50\%, 84\%)\\
& & & & & (deg) & (${\mu}$Hz)\\
\hline
3425851 & 268 & ... & 0.78 & 0.84 & [32.3,\:52.0,\:75.1] & [1.37,\:1.98,\:3.01]\\
3632418 & 975 & 21 & 7.92 & 0.40 & [60.3,\:71.3,\:83.3] & [0.88,\:0.94,\:1.04]\\
4143755 & 281 & 510 & 1.67 & 0.43 & [ 4.6,\:20.7,\:55.4] & [0.09,\:0.32,\:0.91]\\
4349452 & 244 & 25 & 0.94 & 0.91 & [71.3,\:80.6,\:87.1] & [1.41,\:1.49,\:1.59]\\
4914423 & 108 & 103 & 0.63 & 0.42 & [16.0,\:43.4,\:73.2] & [0.34,\:0.62,\:1.44]\\
5094751 & 123 & 109 & 0.54 & 0.40 & [ 3.7,\:15.9,\:49.9] & [0.12,\:0.60,\:2.12]\\
5866724 & 85 & 65 & 1.07 & 0.85 & [66.4,\:75.0,\:84.5] & [1.32,\:1.41,\:1.52]\\
6196457 & 285 & 92 & 1.62 & 1.76 & [ 4.5,\:14.5,\:35.8] & [0.24,\:1.90,\:3.97]\\
6278762 & 3158 & 444 & 3.95 & 2.39 & [48.4,\:64.8,\:78.6] & [0.35,\:0.39,\:0.49]\\
7296438 & 364 & ... & 6.12 & 0.57 & [ 6.8,\:19.2,\:48.7] & [0.15,\:0.50,\:1.11]\\
7670943 & 269 & ... & 0.81 & 0.76 & [66.8,\:77.3,\:86.0] & [1.79,\:1.90,\:2.05]\\
8292840 & 260 & 126 & 1.32 & 0.61 & [64.4,\:75.6,\:85.3] & [1.37,\:1.47,\:1.59]\\
8349582 & 122 & 95 & 1.17 & 0.54 & [10.8,\:34.8,\:66.1] & [0.12,\:0.27,\:0.55]\\
8478994 & 245 & 37 & 1.09 & 1.53 & [19.9,\:38.7,\:64.6] & [0.43,\:0.71,\:1.16]\\
8554498 & 5 & ... & 0.78 & 0.42 & [ 2.3,\:11.9,\:49.6] & [0.07,\:0.44,\:1.67]\\
8866102 & 42 & 410 & 2.28 & 0.66 & [78.4,\:83.6,\:88.0] & [2.03,\:2.07,\:2.12]\\
9414417 & 974 & ... & 3.88 & 0.37 & [46.0,\:58.7,\:76.9] & [0.92,\:1.05,\:1.25]\\
9592705 & 288 & ... & 1.55 & 0.44 & [42.7,\:52.2,\:65.3] & [0.90,\:1.08,\:1.30]\\
10586004 & 275 & 129 & 2.16 & 0.95 & [19.7,\:42.9,\:69.5] & [0.38,\:0.70,\:1.12]\\
10666592 & 2 & 2 & 1.60 & 0.21 & [28.6,\:41.8,\:61.4] & [0.66,\:0.95,\:1.29]\\
11133306 & 276 & 509 & 0.75 & 0.86 & [ 7.6,\:27.1,\:62.4] & [0.21,\:0.61,\:1.53]\\
11295426 & 246 & 68 & 5.86 & 0.32 & [27.6,\:43.1,\:70.2] & [0.21,\:0.30,\:0.46]\\
11853905 & 7 & 4 & 0.98 & 0.42 & [ 8.0,\:36.2,\:66.9] & [0.19,\:0.38,\:0.92]\\
11904151 & 72 & 10 & 1.58 & 0.85 & [ 8.6,\:40.5,\:74.0] & [0.16,\:0.33,\:0.80]\\
\hline
\end{tabular}\\
\end{table*}

\begin{table*}
\caption{Results on {\it Kepler} planet-less stars in category A}
\label{tab:output_pass_nonKOI}
\begin{tabular}{ccccc}
\hline
KIC & HBR$_{\mathrm {max}}$ & $\spl/\Gamma$ & $\is$(16\%, 50\%, 84\%) & $\spl$(16\%, 50\%, 84\%)\\
& & & (deg) & (${\mu}$Hz)\\
\hline
1435467 & 3.35 & 0.52 & [52.7,\:59.5,\:70.4] & [1.46,\:1.64,\:1.82]\\
4914923 & 10.58 & 0.56 & [36.4,\:43.3,\:51.7] & [0.49,\:0.58,\:0.69]\\
5773345 & 5.06 & 0.67 & [27.5,\:31.2,\:35.4] & [1.50,\:1.78,\:2.13]\\
6225718 & 10.21 & 0.70 & [25.1,\:27.2,\:29.8] & [1.52,\:1.70,\:1.85]\\
6679371 & 2.71 & 0.53 & [69.9,\:78.5,\:86.3] & [1.78,\:1.86,\:1.97]\\
7103006 & 2.97 & 0.51 & [41.9,\:46.2,\:51.5] & [1.76,\:1.96,\:2.16]\\
7206837 & 1.80 & 0.93 & [32.3,\:35.4,\:38.8] & [2.46,\:2.74,\:3.03]\\
7510397 & 8.76 & 0.67 & [13.9,\:17.6,\:21.7] & [0.99,\:1.37,\:1.66]\\
7680114 & 5.55 & 0.51 & [15.9,\:24.3,\:52.0] & [0.26,\:0.59,\:0.93]\\
7871531 & 2.91 & 0.60 & [50.7,\:64.1,\:81.2] & [0.34,\:0.40,\:0.47]\\
7970740 & 4.66 & 0.57 & [40.2,\:48.5,\:61.5] & [0.29,\:0.37,\:0.46]\\
8006161 & 8.78 & 0.62 & [33.1,\:39.6,\:47.5] & [0.45,\:0.54,\:0.64]\\
8179536 & 2.37 & 0.71 & [47.8,\:53.3,\:60.3] & [1.55,\:1.72,\:1.91]\\
8379927 & 8.21 & 0.54 & [67.0,\:71.9,\:78.8] & [1.10,\:1.15,\:1.20]\\
8394589 & 3.33 & 0.70 & [63.7,\:69.1,\:76.0] & [1.00,\:1.06,\:1.11]\\
9025370 & 2.75 & 0.60 & [52.5,\:69.9,\:83.6] & [0.41,\:0.47,\:0.55]\\
9139151 & 3.37 & 0.57 & [61.0,\:71.2,\:83.0] & [0.93,\:1.00,\:1.09]\\
9139163 & 2.88 & 0.83 & [23.1,\:25.1,\:27.6] & [3.09,\:3.50,\:3.85]\\
9965715 & 3.10 & 0.62 & [55.5,\:60.3,\:66.2] & [1.83,\:1.96,\:2.09]\\
11253226 & 2.31 & 0.53 & [45.8,\:48.3,\:50.9] & [3.10,\:3.24,\:3.39]\\
12009504 & 3.83 & 0.53 & [63.6,\:69.2,\:77.1] & [1.12,\:1.20,\:1.27]\\
12069424 & 35.68 & 0.54 & [45.9,\:51.9,\:60.2] & [0.45,\:0.50,\:0.55]\\
\hline
\end{tabular}\\
\end{table*}

\begin{table*}
\caption{Results on {\it Kepler} planet-less stars in category B}
\label{tab:output_fail_nonKOI}
\begin{tabular}{ccccc}
\hline
KIC & HBR$_{\mathrm {max}}$ & $\spl/\Gamma$ & $\is$(16\%, 50\%, 84\%) & $\spl$(16\%, 50\%, 84\%)\\
& & & (deg) & (${\mu}$Hz)\\
\hline
2837475 & 1.74 & 0.46 & [70.9,\:76.6,\:83.6] & [3.01,\:3.10,\:3.20]\\
3427720 & 3.18 & 0.31 & [12.4,\:28.5,\:61.0] & [0.19,\:0.43,\:0.88]\\
3456181 & 3.52 & 0.28 & [34.5,\:48.6,\:72.0] & [0.75,\:0.99,\:1.36]\\
3656476 & 11.57 & 0.42 & [34.6,\:48.8,\:73.1] & [0.24,\:0.31,\:0.42]\\
3735871 & 1.42 & 0.38 & [42.9,\:66.2,\:83.1] & [0.62,\:0.72,\:0.98]\\
5184732 & 13.74 & 0.43 & [58.8,\:70.8,\:83.0] & [0.53,\:0.57,\:0.63]\\
5950854 & 2.16 & 1.04 & [ 1.7,\: 8.2,\:27.6] & [0.05,\:1.15,\:2.35]\\
6106415 & 19.61 & 0.44 & [67.0,\:75.9,\:85.1] & [0.68,\:0.71,\:0.75]\\
6116048 & 12.58 & 0.42 & [62.8,\:73.0,\:83.7] & [0.61,\:0.64,\:0.70]\\
6508366 & 2.56 & 0.49 & [80.9,\:85.6,\:88.7] & [2.12,\:2.19,\:2.26]\\
6603624 & 17.93 & 0.44 & [ 2.0,\: 4.2,\:38.7] & [2.00,\:0.30,\:1.44]\\
6933899 & 10.51 & 0.31 & [48.9,\:64.2,\:81.0] & [0.33,\:0.37,\:0.45]\\
7106245 & 1.65 & 0.43 & [13.5,\:28.6,\:62.6] & [0.27,\:0.57,\:1.36]\\
7771282 & 1.17 & 0.39 & [48.6,\:67.0,\:82.0] & [1.05,\:1.19,\:1.39]\\
7940546 & 9.25 & 0.35 & [52.5,\:63.0,\:76.6] & [0.97,\:1.08,\:1.23]\\
8150065 & 1.39 & 0.28 & [ 5.2,\:26.0,\:65.8] & [0.15,\:0.49,\:1.37]\\
8228742 & 6.52 & 0.44 & [29.9,\:38.5,\:58.6] & [0.56,\:0.83,\:1.11]\\
8424992 & 2.72 & 0.41 & [ 4.0,\:19.2,\:59.3] & [0.08,\:0.31,\:1.04]\\
8694723 & 7.59 & 0.46 & [32.4,\:37.4,\:43.0] & [1.10,\:1.25,\:1.46]\\
8760414 & 7.19 & 0.43 & [ 2.1,\: 8.1,\:40.4] & [0.04,\:0.48,\:1.74]\\
8938364 & 8.91 & 0.31 & [ 7.8,\:25.0,\:61.7] & [0.10,\:0.23,\:0.65]\\
9098294 & 3.86 & 0.36 & [30.8,\:49.9,\:75.1] & [0.33,\:0.43,\:0.66]\\
9206432 & 1.62 & 0.30 & [21.1,\:36.2,\:59.6] & [1.06,\:1.73,\:2.77]\\
9353712 & 1.78 & 0.87 & [21.0,\:28.9,\:53.8] & [0.86,\:1.84,\:2.65]\\
9410862 & 1.66 & 0.78 & [13.7,\:21.2,\:45.3] & [0.47,\:1.16,\:2.01]\\
9812850 & 1.82 & 0.38 & [50.7,\:64.7,\:81.3] & [1.40,\:1.57,\:1.87]\\
10068307 & 11.91 & 0.40 & [33.4,\:41.9,\:58.4] & [0.58,\:0.77,\:0.96]\\
10079226 & 2.23 & 2.50 & [49.8,\:71.6,\:84.0] & [0.64,\:0.75,\:0.93]\\
10162436 & 6.61 & 0.29 & [28.7,\:44.6,\:62.2] & [0.48,\:0.65,\:1.02]\\
10454113 & 3.23 & 0.42 & [26.4,\:34.5,\:47.6] & [0.83,\:1.16,\:1.53]\\
10516096 & 6.79 & 0.33 & [53.2,\:70.2,\:83.8] & [0.45,\:0.49,\:0.58]\\
10644253 & 1.86 & 0.21 & [ 2.5,\:13.9,\:54.4] & [0.06,\:0.34,\:1.36]\\
10730618 & 1.01 & 0.31 & [11.4,\:24.6,\:52.8] & [0.52,\:1.32,\:2.65]\\
11081729 & 1.32 & 2.30 & [80.7,\:85.4,\:88.6] & [3.22,\:3.40,\:3.51]\\
11772920 & 1.93 & 0.49 & [51.1,\:67.0,\:81.8] & [0.28,\:0.33,\:0.40]\\
12069127 & 1.66 & 0.19 & [16.9,\:40.4,\:70.3] & [0.35,\:0.65,\:1.16]\\
12069449 & 29.29 & 0.37 & [33.1,\:47.0,\:70.7] & [0.27,\:0.35,\:0.49]\\
12258514 & 13.77 & 0.30 & [19.9,\:34.0,\:64.6] & [0.28,\:0.46,\:0.81]\\
12317678 & 2.74 & 0.20 & [46.1,\:62.4,\:80.5] & [0.92,\:1.06,\:1.34]\\
\hline
\end{tabular}\\
\end{table*}

We perform the asteroseismic analysis using the MCMC method, and
summarize the main results in Tables
\ref{tab:output_pass_KOI},\ref{tab:output_fail_KOI} and
\ref{tab:output_pass_nonKOI},\ref{tab:output_fail_nonKOI} for stars with
and without planets, respectively.  We classify those stars as category
A if their measured values of ($\is, \spl/\Gamma$) satisfy the three
analytic criteria (I), (II), and (III) with $\alpha=10$ and $\beta=15$.
Otherwise the stars are classified as category B.  There are 9 stars with planets of category
A, and 22 without planets. The
classification is admittedly not strict, because it is based on the
measured median values neglecting the quoted errors, in addition to the
qualitative nature of the criteria themselves.  Nevertheless such a
classification is useful as a rough measure of the reliability of the
inference.

Upper and lower panels in Figure \ref{fig:KOI_scatter} plot the
distribution of measured $\is$ and $\spl/\Gamma$ for stars with and
without planets, respectively. Stars belonging to categories A and B are
plotted in filled circles with error-bars and in crosses,
respectively. Also we indicate the KOI number for category A stars with
planets in the upper panel. Since the target selection is somewhat
heterogeneous, we cannot put any strong conclusion at this
point. Nevertheless it is interesting to note that the category A stars
with planets are preferentially located around the large $\is$ region
relative to those without planets, suggesting a spin-orbit alignment of
transiting planets in general. Further implications of the results for
the {\it Kepler} will be described in our next paper (Kamiaka, Benomar,
and Suto, in preparation).

\begin{figure*}
\includegraphics[width=\columnwidth]{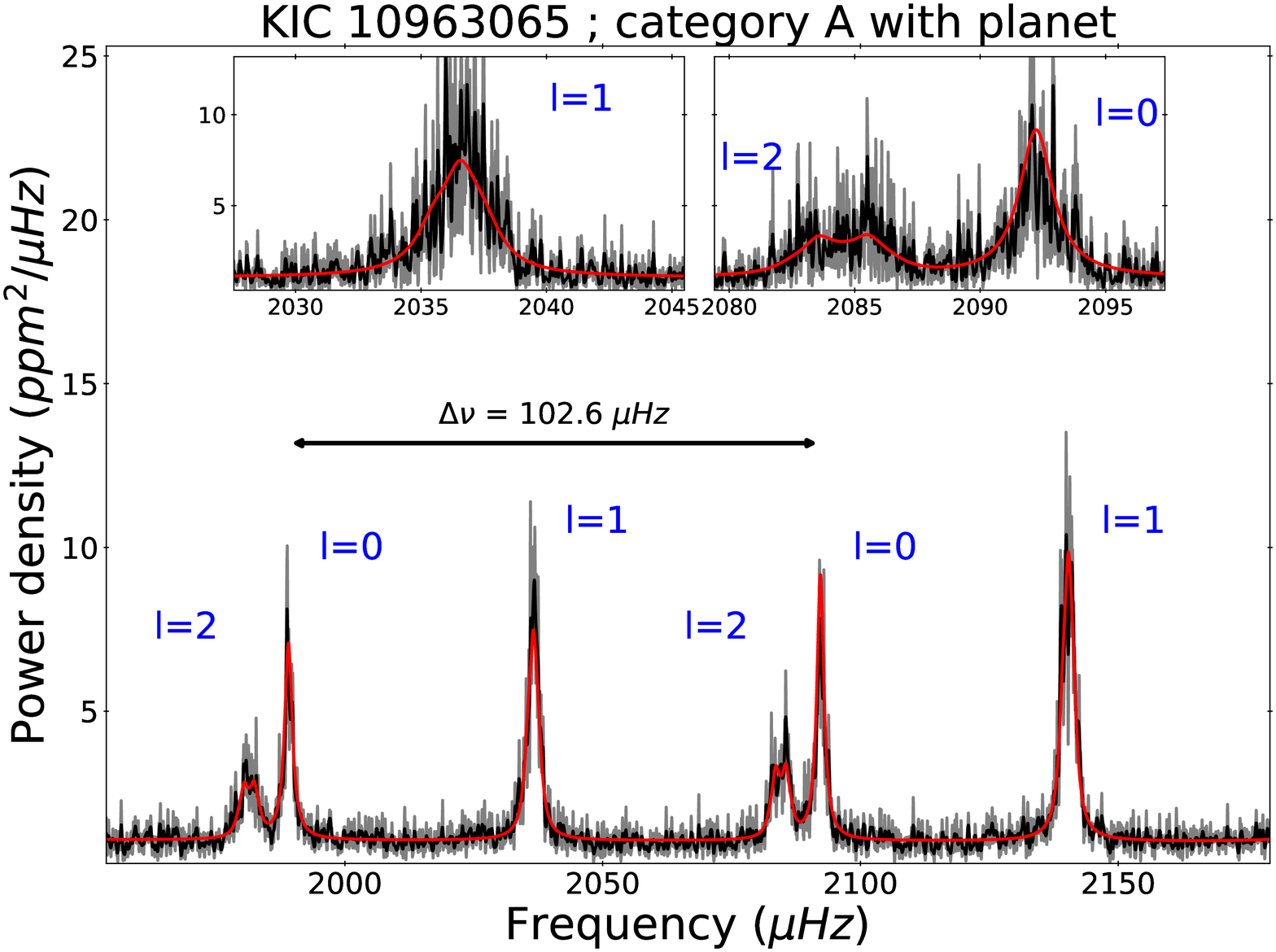}
\includegraphics[width=\columnwidth]{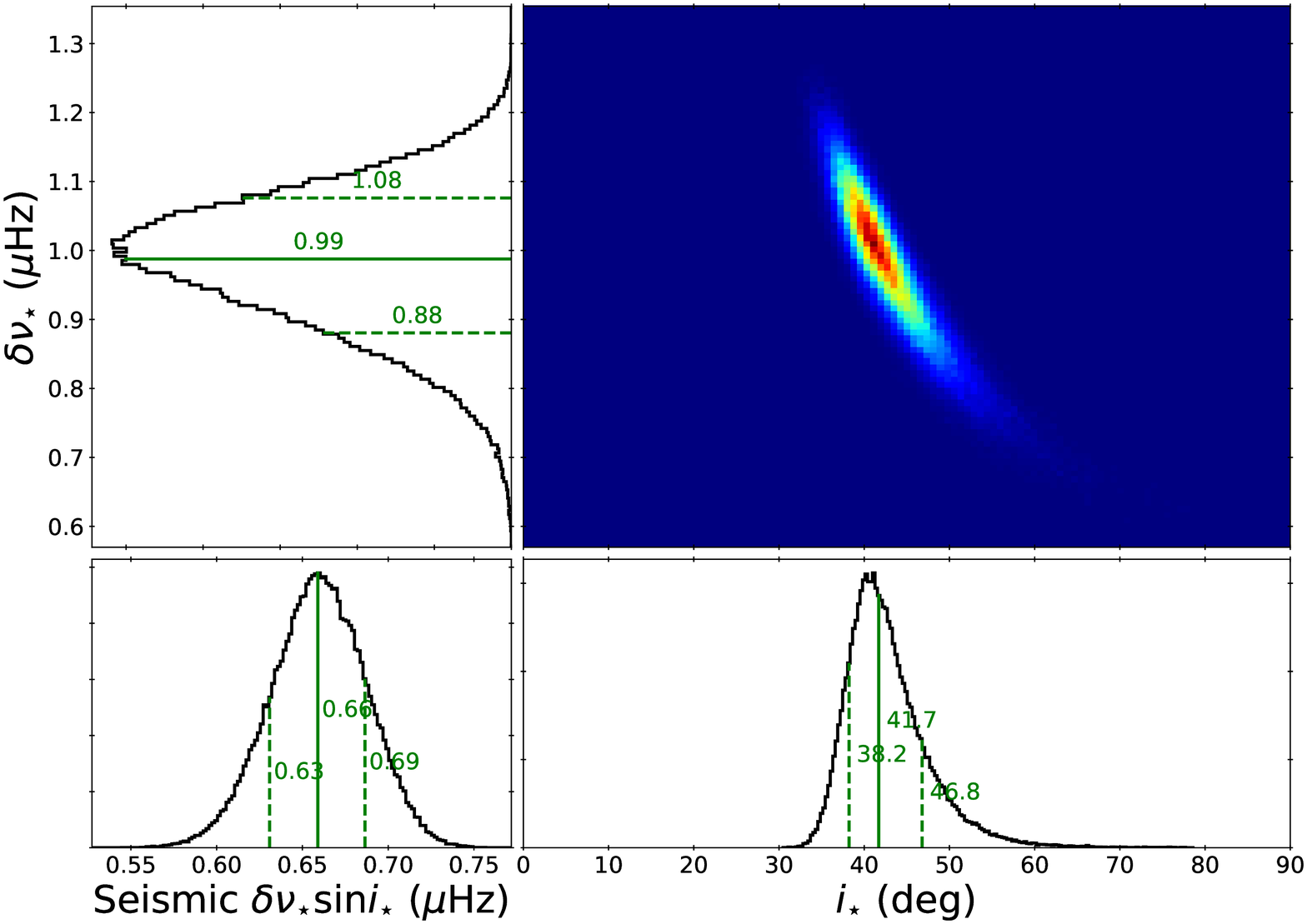}
\caption{
Results of our asteroseismic analysis for KIC 10963065, as a
typical example of planet-host stars in category A.  Left: power spectra
around $\numax$ along with the best-fit model curve (red solid line).
Black and gray lines indicate spectra Gaussian-smoothed over the width
of 0.15$\dnu$ and 0.05$\dnu$, respectively.  The upper insets display the zoom-in
views of the spectra, around $l=1$ (left) and $l=2$ and $0$ (right)
modes.  Black and gray lines in these insets are Gaussian-smoothed over
0.03$\dnu$ and 0.01$\dnu$.  Right: upper-right panel is the
two-dimensional correlation map of $\is$ and $\spl$ from the MCMC
sampling.  Top-left and bottom-right panels are the corresponding
marginalized PPD for $\spl$ and $\is$, respectively.  Bottom-left panel
is the PPD of seismic $\spl\sin{\is}$.  Green solid and dashed lines in
these histograms indicate the median and 1$\sigma$ credible regions.
}
\label{fig:10963065}
\end{figure*}

\begin{figure*}
\includegraphics[width=\columnwidth]{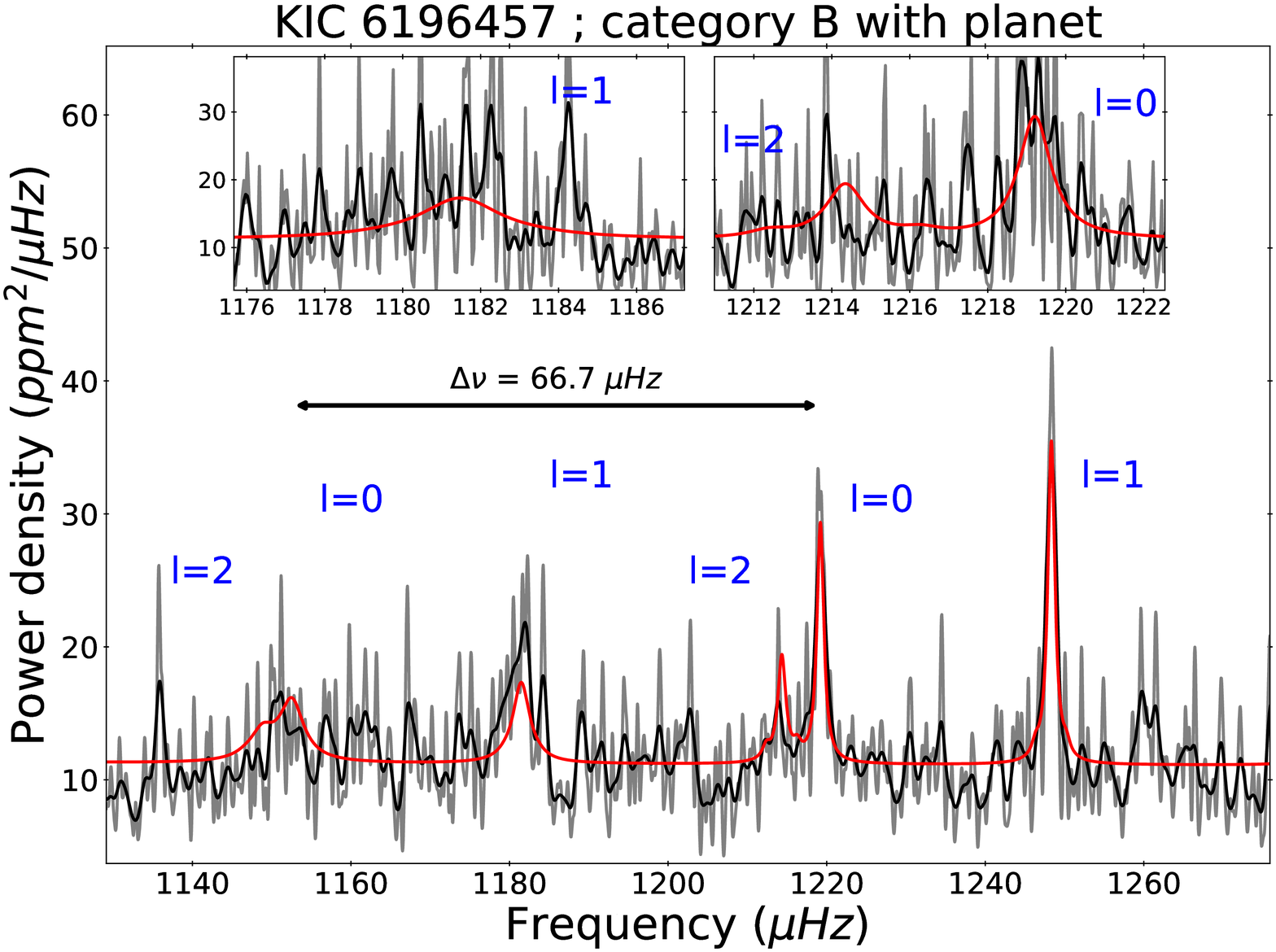}
\includegraphics[width=\columnwidth]{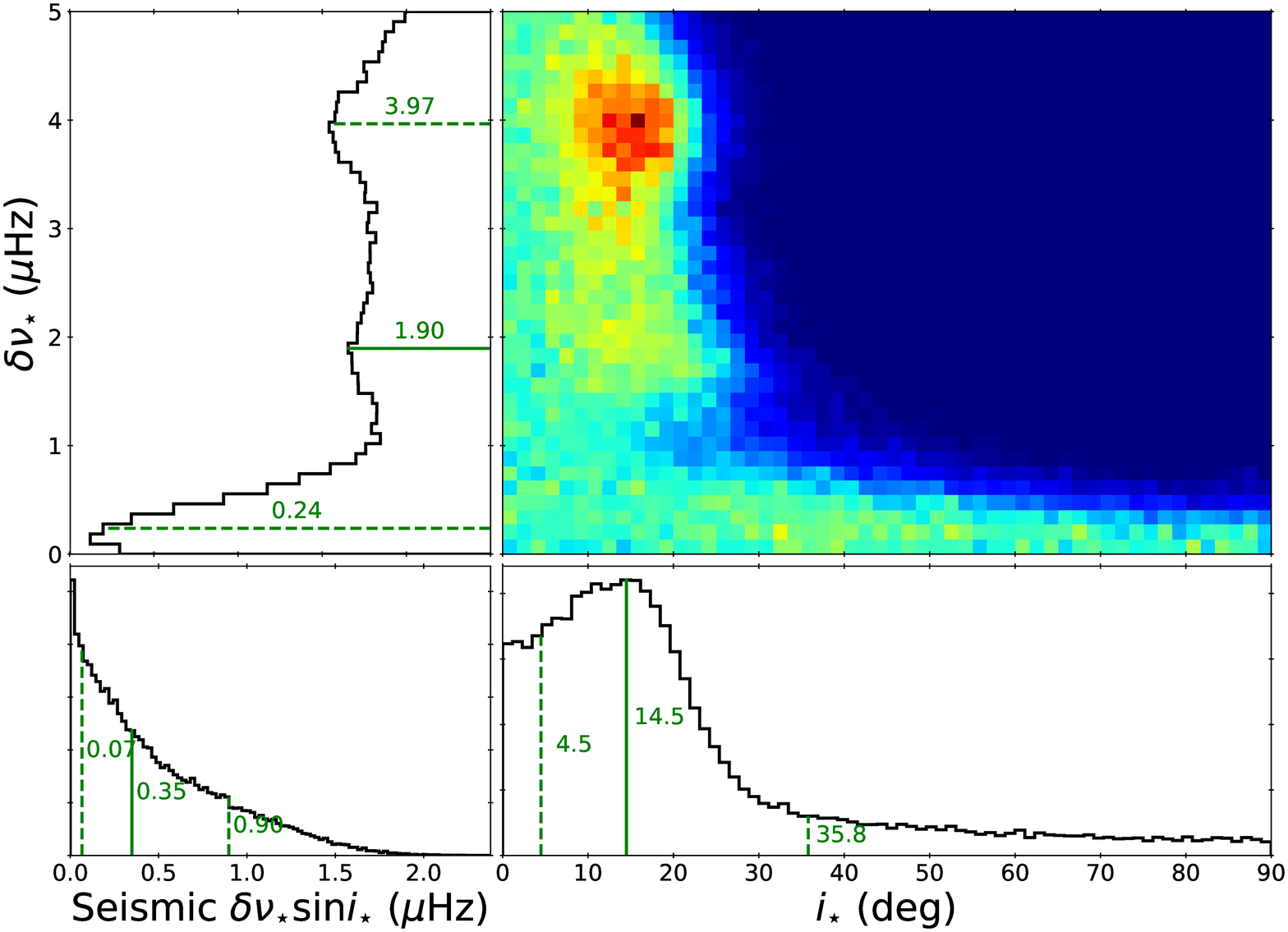}
\caption{
Same as Figure \ref{fig:10963065}, but for KIC 6196457.
}
\label{fig:6196457}
\end{figure*}

Figures \ref{fig:10963065} and \ref{fig:6196457} show examples of
power spectra and the resulting two dimensional PPD of $\is$ and
$\spl$ for categories A and B stars.  They present the difference of
the constraining power on $\is$ and $\spl$ between our categories A
and B. Note, however, that these two may be extreme examples, and in
some cases the difference between A and B is
milder.

\subsection{Consistency of asteroseismically-derived parameters with other observations
\label{subsec:comparison}}

\begin{figure*}
\includegraphics[width=0.95\columnwidth]{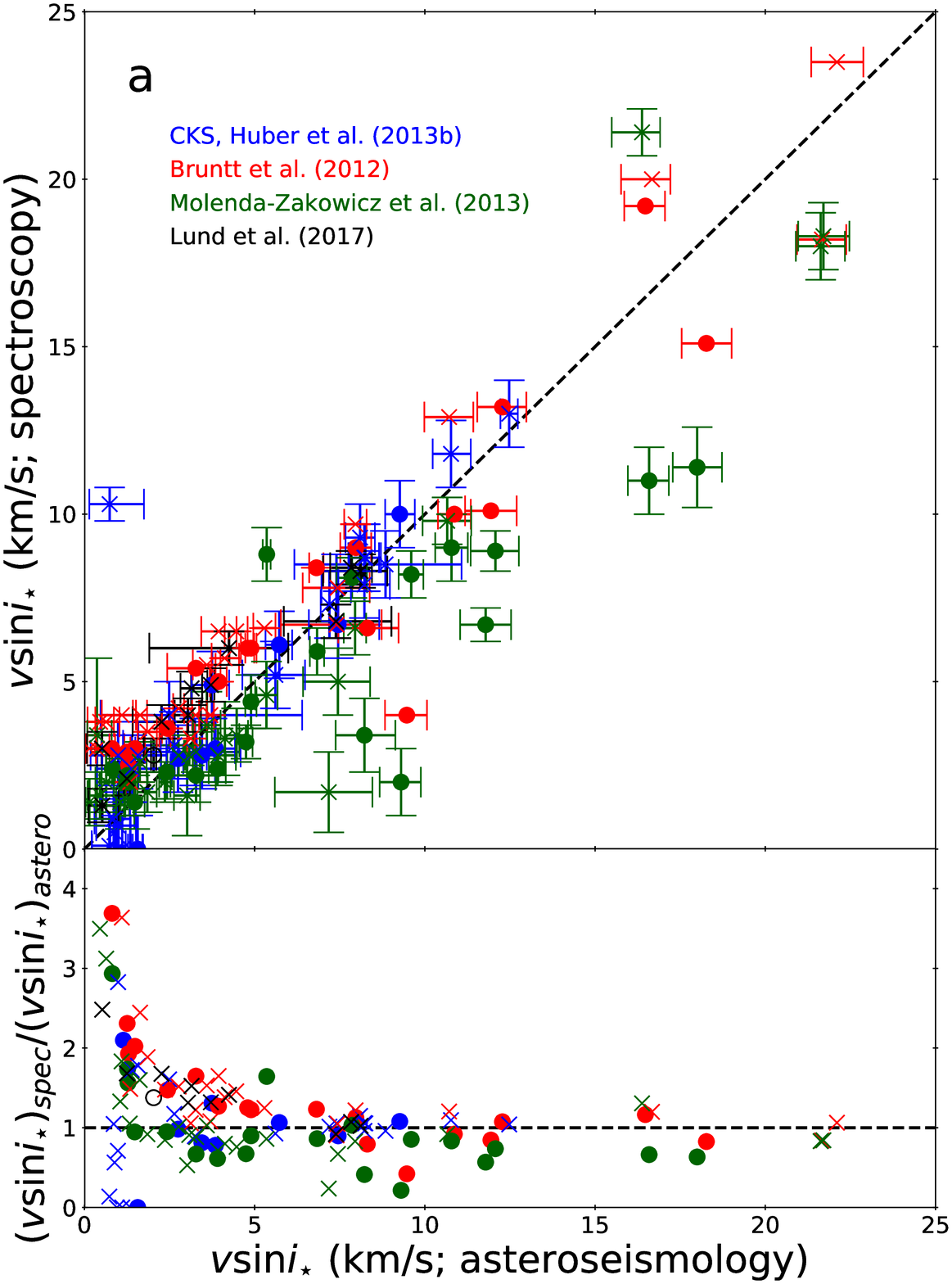}
\includegraphics[width=0.95\columnwidth]{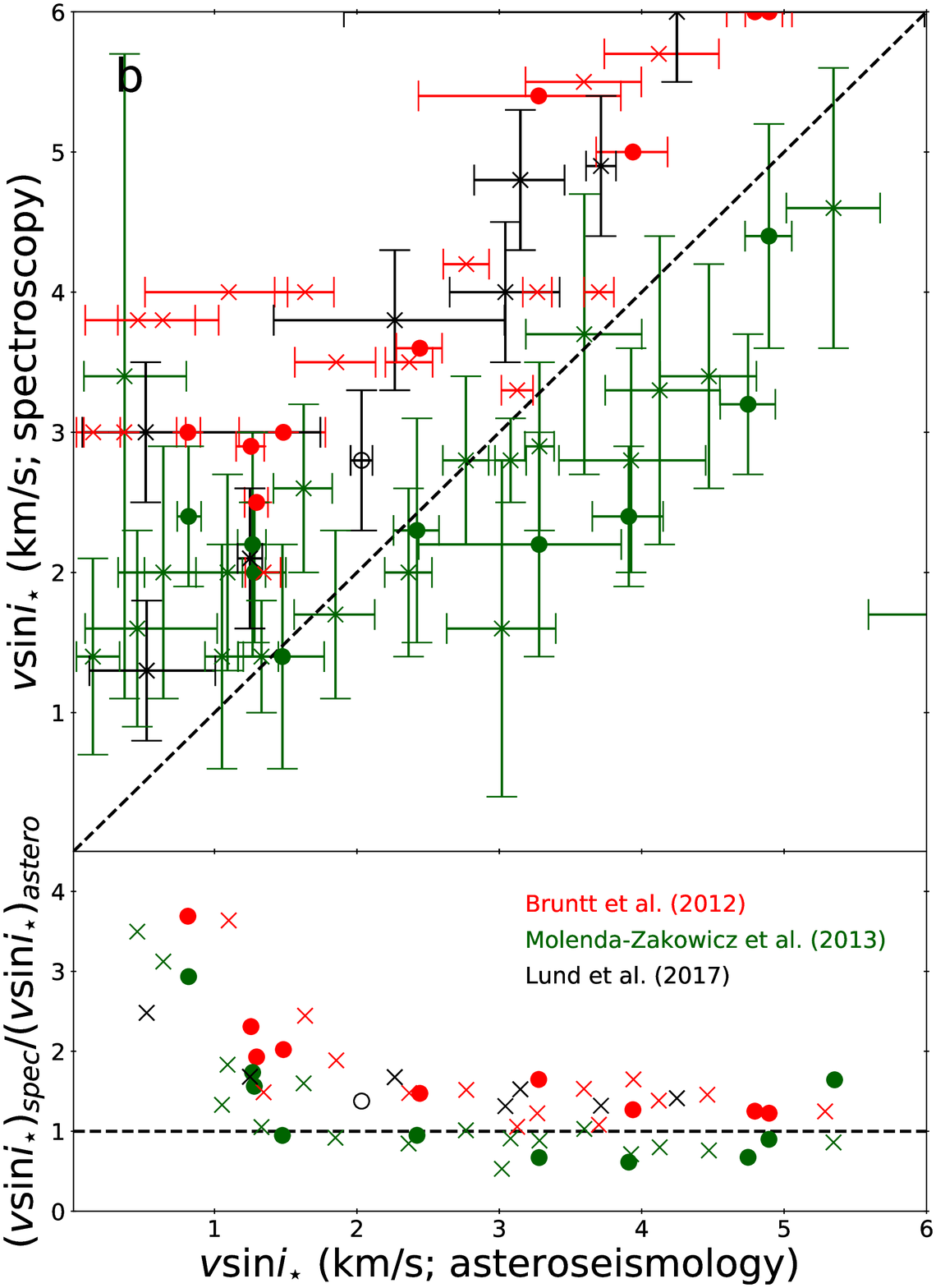}
\caption{
Comparison of $\vsini$ estimated from spectroscopy and those
from asteroseismology. Filled circles and crosses correspond to
categories A and B stars, respectively. Figure a (left) shows
stars with planets (blue), with spectroscopic values from CKS (32
stars) and \citealt{Huber:2013ab} (1 star). Shown stars without
planets use values from \citealt{Lund:2017aa} in black (11 stars),
from \citealt{Bruntt:2012aa} in red (43 stars), and from
\citealt{Molenda-Zakowicz:2013aa} in green (46 stars). Figure b
(right) is an enlarged view for stars without planet whose
$\vsini$ is less than 6km/s.
}
\label{fig:comp_vsini}
\end{figure*}

\begin{figure}
\includegraphics[width=1.\columnwidth]{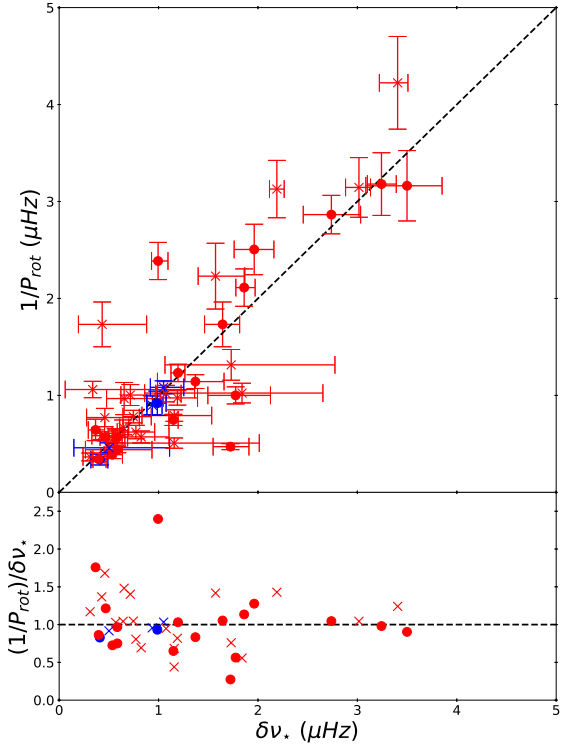}
\caption{
Comparison of $\spl$ derived from asteroseismology and the
inverse of the stellar rotation period derived from photometric
variations, whenever available. Stars with and without planets are plotted in
blue and red, respectively. Filled circles and crosses correspond to
categories A and B stars.
}
\label{fig:comp_spl}
\end{figure}

\begin{figure*}
\includegraphics[width=0.95\columnwidth]{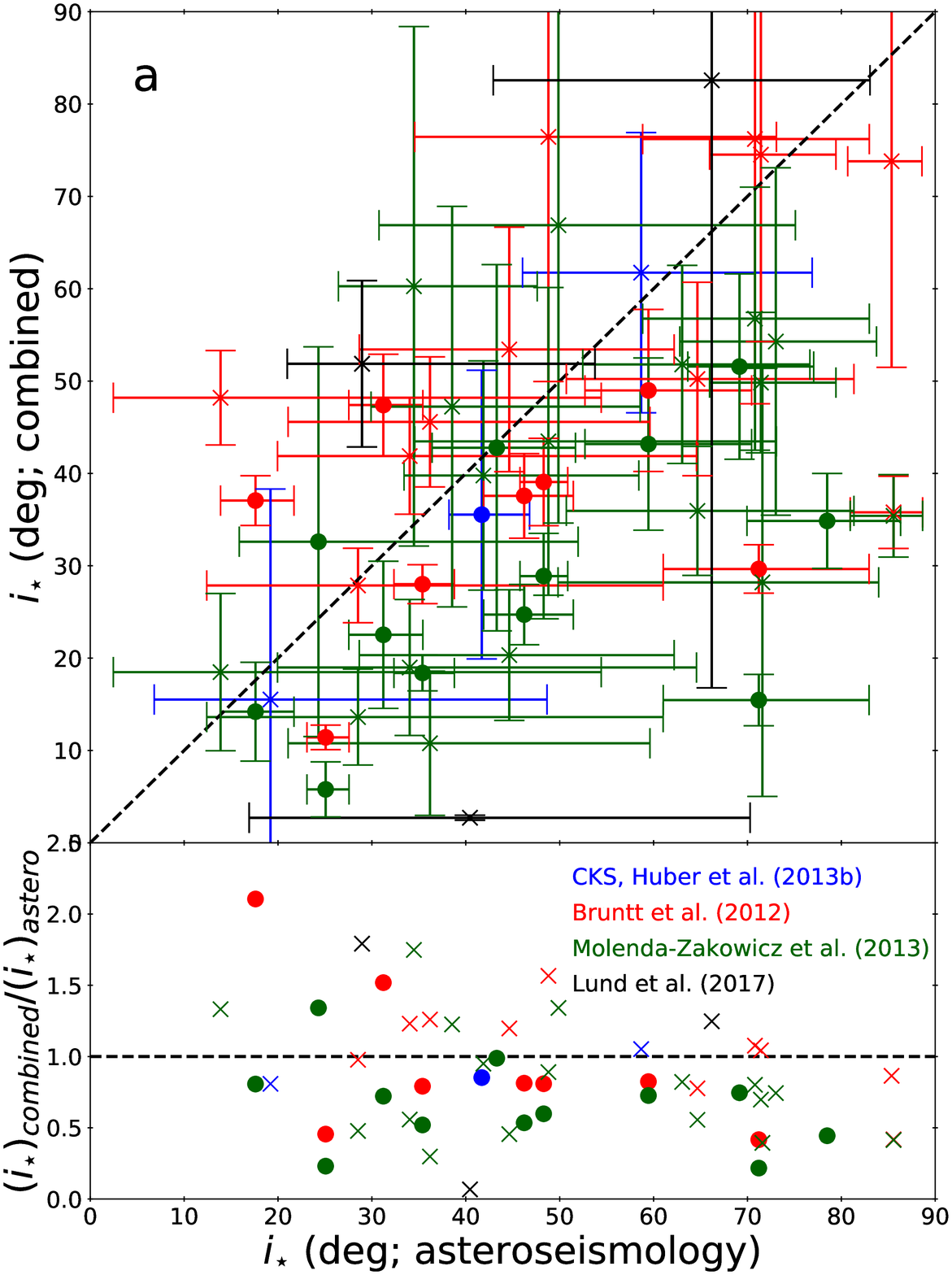}
\includegraphics[width=0.95\columnwidth]{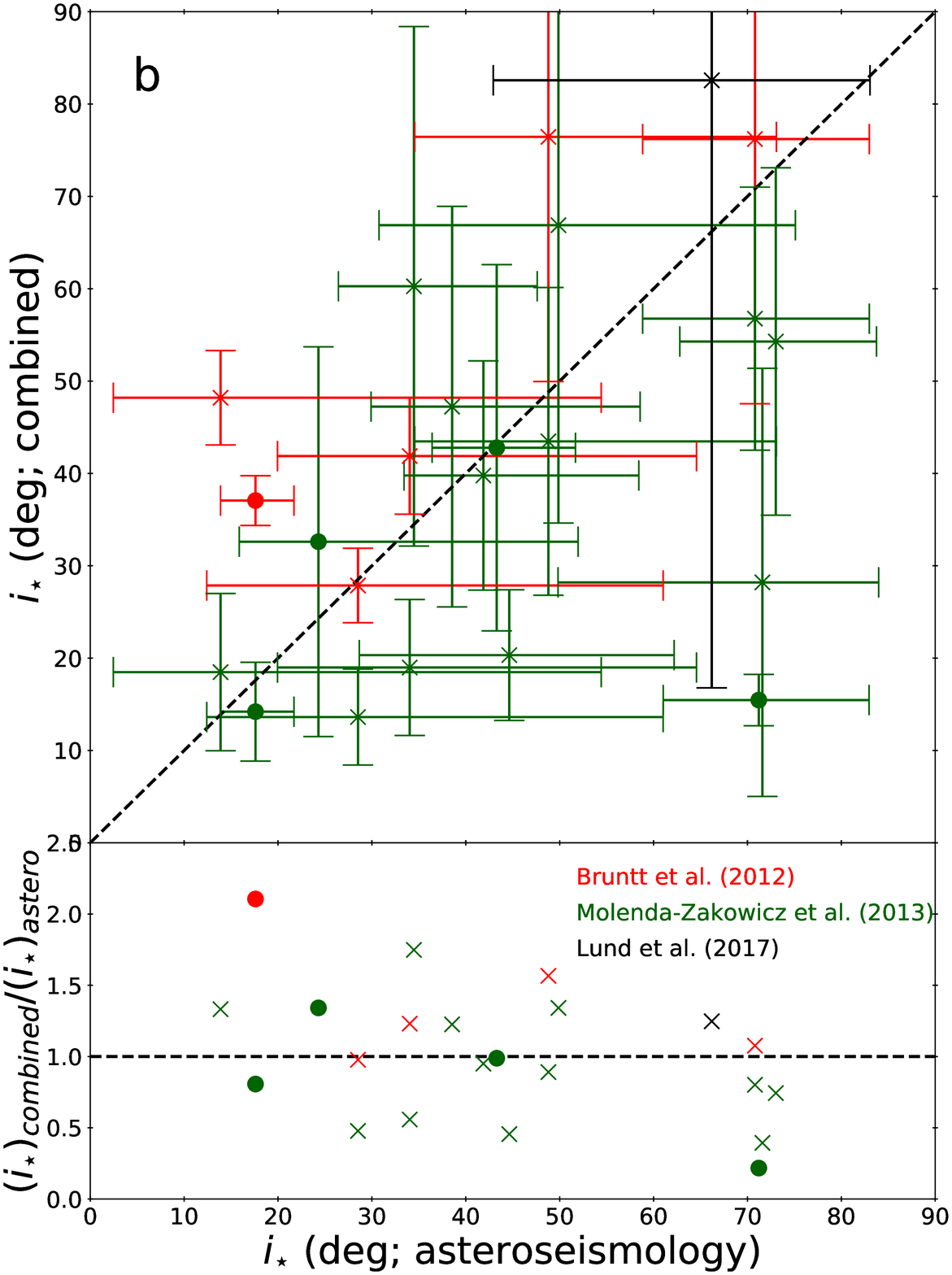}
\caption{ 
Comparison of $\is$ derived from asteroseismology and those derived
from spectroscopic and time-domain photometric observations
(equation \ref{eq:is_combined}).
The colors and symbols are the same as in Figure \ref{fig:comp_vsini}.
}
\label{fig:comp_inc}
\end{figure*}

Unlike for simulated stars, the true values of the stellar parameters for actual {\it Kepler}
stars are obviously not known.  Thus it is important to compare our asteroseismic
estimates of stellar parameters against other independent observations,
which is attempted in this subsection.

We first consider $\vsini$ that can be measured also from line widths of
spectroscopically observed stars. For 33 stars with planets, we use the
spectroscopic $\vsini$ from California-Kepler Survey (CKS;
https://california-planet-search.github.io/cks-website/), except one
from \citet{Huber:2013ab}.

For 61 stars without planets, we consider two different spectroscopic
datasets from \citet{Bruntt:2012aa} and
\citet{Molenda-Zakowicz:2013aa}.  First we adopt the data for 43 stars
from \citet{Bruntt:2012aa}. Out of the remaining 18 stars not listed
in their catalog, we adopt the data from \citet{Lund:2017aa} for 11
stars.  Next we repeat the same procedure starting with the dataset of
\citet{Molenda-Zakowicz:2013aa}. In this case, we combine 46 stars
from \citet{Molenda-Zakowicz:2013aa}, and 11 stars from
\citet{Lund:2017aa}.

Note that $\vsini$ in \citet{Huber:2013ab} and \citet{Lund:2017aa} is
calculated on the basis of the Stellar Parameter Classification
pipeline (SPC; \citealt{Buchhave:2012aa}), while CKS,
\citet{Bruntt:2012aa}, and \citet{Molenda-Zakowicz:2013aa} develop
their own pipeline in computing $\vsini$.

While $\vsini$ can be directly estimated from
spectroscopic data, it is not the case for asteroseismolgy.
We estimate the stellar radius $R_{\star}$ from 
the scaling relation calibrated with the Sun:
\begin{eqnarray}
\frac{R_{\star}}{R_{\odot}} = 
\left(\frac{\nu_{\mathrm{max},\star}}{\numaxsun}\right)
\left(\frac{\dnu_\star}{\dnusun}\right)^{-2}
\left(\frac{T_{\rm eff, \star}}{\Teffsun}\right)^{1/2},
\end{eqnarray}
where $\nu_{{\rm max},\star}$ is the frequency corresponds to the peak
of the mode heights (see Figure \ref{fig:16CygA_profile}),
$\dnu_\star$ is the large separation, and $T_{\rm eff, \star}$ is the
effective temperature of the star.  Thus $R_\star$ can be estimated
from the two asteroseismic observables, $\numax$ and $\dnu_{\star}$,
along with $T_{\rm eff, \star}$, which leads to asteroseismic estimate
of $\vsini$:
\begin{eqnarray}
\vsini\:({\rm asteroseismology}) = 2{\pi}R_{\star}\spl\sin{\is}.
\end{eqnarray}
We adopt $\nu_{\mathrm{max},\odot}=3100\,\mu$Hz, $\dnu_{\odot} =
134.9\,\mu$Hz, and $\mathrm{T}_{\mathrm{eff},\odot}=5777\:$K
\citep{Broomhall:2009aa,Gaulme:2016ab}.

Figure \ref{fig:comp_vsini} compares $\vsini$ from asteroseismic and
spectroscopic data. The left panel shows planet-hosting stars in
blue (32 from CKS and 1 from \citealt{Huber:2013ab}), and also stars
without planets, 11 from \citealt{Lund:2017aa} in black, 43 from
\citealt{Bruntt:2012aa} in red, and 46 from
\citealt{Molenda-Zakowicz:2013aa} in green.  A significant fraction
of stars without planets overlaps in the two sources, and thus we
distinguish them using different colors.  Filled circles and crosses
correspond to categories A and B stars, respectively. We do not
quote error-bars for \citet{Bruntt:2012aa} since they are not
available from the published table.

The left panel of Figure \ref{fig:comp_vsini} suggests that
asteroseismic and spectroscopic $\vsini$ are in reasonable
agreement.  However, a closer look at $\vsini < 6$km/s data in the
right panel reveals an interesting feature; the estimates by
\citet{Bruntt:2012aa} (red) are systematically larger than our
asteroseismic values, while those by \citet{Molenda-Zakowicz:2013aa}
(green) are systematically smaller. Our result are somewhere
in-between, except for $\vsini \lesssim 2$km/s.  Since these authors
have a large fraction of stars in common, the feature should not be
due to differences in the stellar properties.  We suspect that the
difference between the two spectroscopic results comes from the
subtle modeling of micro/macro-turbulence effects in
spectroscopic data.  We would like to point that 
``the roundest A-type star'' KIC 11145123
\citep{Kurtz2014,Gizon2016} presents an interesting example
in this context. \citep{Takada2017} found that
the spectroscopically measured value of $\vsini \simeq 5$km/s
suffers from systematic overestimate, and asteroseismically derived
equatorial rotation velocity of $\vsini \simeq 1$km/s proved to be
more reliable. This suggests that the spectroscopic
measurement of $\vsini$ for slowly rotating stars needs to be
interpreted with caution, which is in good agreement with our
conclusions from Figure \ref{fig:comp_vsini}.
The importance of the careful
calibration of turbulence has been well recognized in earlier publications, for instance by \citet{Bruntt:2012aa}. The lower panels of Figure \ref{fig:comp_vsini} provide observational evidences of this problem.
Incidentally, the
overall consistency between asteroseismic and spectroscopic $\vsini$
($>$ a few km/s) may also reinforce the nearly-uniform rotation of
stars as stated by \citet{Benomar2015}.  This is because
asteroseismology measures the stellar rotation averaged over its
interior, while spectroscopy measures its surface rotation.

Figure \ref{fig:comp_spl} compares asteroseismic $\spl$ and the inverse
of stellar rotation period measured from photometric variability for 46
stars \citep{Garcia:2014aa}. While they agree reasonably on average,
individual agreement is not good except for $\spl \gtrsim 2 \mu$Hz. Again
both the photometric variation and rotational splitting are not reliably
identified for slowly rotating stars.

Combining the spectroscopic $\vsini$, asteroseismic $R_{\star}$, and
photometric $P_{\rm rot}$, we can estimate $\is$ as
\begin{eqnarray}
\label{eq:is_combined}
\is\:({\rm combined}) = \sin^{-1} \left(
\frac{P_{\rm rot}}{2{\pi}R_{\star}} \vsini
\right).
\end{eqnarray}
Figure \ref{fig:comp_inc} is similar to Figure \ref{fig:comp_vsini},
but instead, compares $\is$ estimated from equation
(\ref{eq:is_combined}) with the asteroseismic $\is$.
Panel a (left) shows the stars whose $\is$ is derived from the combined analysis, while panel b (right) shows planet-less stars with $v\sin{\is} < 6$km/s alone, similarly to Figure \ref{fig:comp_vsini}.
The large scatter, that is mainly due to the
photometric variation uncertainty, makes it difficult to draw any definite
conclusion at this point.
Indeed, the lightcurve modulation attributed to spots could be affected by the fact that the number, lifetime and latitude of the spots are unknown. It is therefore difficult to identify the reason of the scatter at this stage. However, this is the current status of
the mutual comparison of independently measured $\is$, which needs to
be kept in mind in considering the implications of the distribution of
$\is$.

This caution may be relevant to interpret the recent results by
\cite{Kovacs2018arXiv}, which reports the possible alignment of
stellar inclinations in the Praesepe cluster from the combined
analysis of photometric rotation periods, spectroscopic rotation
velocities, and estimated stellar radii. Figures 3 and 4 of
\cite{Kovacs2018arXiv} indicate that the cumulative distribution of
$\is$ for the cluster is biased toward the larger value relative to
the isotropic distribution. This could be explained as well if the
macro-turbulence is underestimated as it might be the case for
\citet{Bruntt:2012aa}. It is premature to firmly conclude at this
point, but it clearly indicates the importance of our current
finding exhibited in Figure \ref{fig:comp_vsini}, and the necessity
to perform in the future a thorough comparative study of methods for
inferring the stellar inclination.

\section{Discussion}
\label{sec:discussion}

One of our main findings is that the seismology provides reliable
stellar inclination only for stars with $20^\circ \lesssim \is \lesssim 80^\circ$, $\spl/\Gamma \gtrsim 0.5$, with high signal-to-noise ratio, and with longer observations. A significant bias arise when this is not the case, so that the stellar inclination could be overestimated for low inclinations and underestimated otherwise. Below we discuss
more broadly its implication on previous results.

\subsection{Inclinations on CoRoT stars}

Although the statistics is low, it is interesting to note that
the analysis of solar-like stars observed by CoRoT
\citep{Baglin2006b, Baglin2006a} often led to low and medium stellar
inclinations.  An isotropic distribution of spins in the sky should
give instead a larger proportion of stars with high inclination.  We
have $\is = 45^\circ \pm 4^\circ$ for HD 181420 \citep{Barban2009},
$\is = 24^\circ \pm 3^\circ$ for HD 181906 \citep{Garcia2009}, and
they are based on the low SNR.  On the other hand, we have $\is
= 17^\circ \pm 9^\circ$ or $\is = 26^{17}_{7}\,^\circ$ for HD 49933
\citep{Benomar:2009ab, Benomar2015} and $\is = 71^\circ \pm
6^\circ$ for HD 49385 \citep{Deheuvels:2010aa} with high SNR.

Those CoRoT stars were observed only for 90 to 180 days, with a
signal-to-noise that does not exceed $\sim 5$. From Figure
\ref{fig:bias}, we expect a substantial bias toward lower
inclinations for most of the CoRoT stars, in agreement with the
apparent excess of low to medium stellar inclinations. This is also
largely consistent with our {\it Kepler} data analysis plotted in Figure
\ref{fig:KOI_scatter}, especially for stars without planet in which
the correlation with the transiting planetary orbital plane should not
exist.

\subsection{Implication for evolved stars
\label{subsec:implications}}

While the current work is specifically dedicated to low-mass
main-sequence stars, our results can be of importance also for evolved
stars. Subgiants and red-giant stars show mixed modes, arising from
the coupling between pressure modes and gravity modes.  Mixed modes
can be mostly sensitive either to the envelope (pressure-like modes)
or to the interior (gravity-like modes). The large number of $l=1$
mixed modes observed in evolved stars has enabled detailed studies of
the interior and evolution of those stars
\citep[e.g.,][]{Deheuvels2012, Mosser2014}.

Because pressure-like modes are short-lived \footnote{The lifetime of
the modes is inversely proportional to the mode width.} and probe
the slowly rotating envelope (e.g., rotations of the order $\sim 100$
days in red-giants), it is expected that split-components of
$l=1$ modes suffer from a severe blending ($\spl \ll \Gamma$). On
the contrary, gravity-like modes have lifetimes of the order of
years and probe regions that mostly rotate faster than the envelope
\citep{Beck2012Nature, Deheuvels2012, Benomar2013a, Deheuvels2014,
Deheuvels2015, Mosser2017}, so that split-components are well
separated ($\spl \gg \Gamma$).

In these conditions and as suggested by Figure \ref{fig:bias}, it is
likely that gravity-like modes allow an accurate determination of the
stellar inclination, provided that they have a significant signal-to-background ratio.
However, one needs to be cautious when determining
the stellar inclination from pressure-like modes. We also stress that
when modes of evolved stars are fitted individually (e.g., using a
local fit, rather than a global fit as performed in this study),
inclinations of blended modes or low signal-to-background modes are
expected to be significantly biased towards lower values.

This suggest that the seismic determination of the stellar inclination
for the red-giant Kepler-56 \citep{Huber:2013aa}, reported to have a
large $\is$ and to host multiple transiting planets, remains certainly
accurate because the analysed split component of the $l=1$ modes are
clearly well resolved and of high signal-to-noise ratio (see their
Figure 1).  However, results from \citet{Corsaro2017Nature} on spin
alignment of star clusters may require a careful interpretation
because they fit different modes independently and determine
\emph{a posteriori} the stellar inclination. In addition, the clusters
consist of faint stars with modes of relatively low amplitudes. As
suggested by Figure \ref{fig:bias}, this may bias stellar inclinations
towards $\sim 30$ degrees.  This indicates the importance of studying a
potential bias on stellar inclination for subgiants and red-giants as
we have performed for main-sequence stars.

\section{Summary}
\label{sec:summary}

The measurement of the stellar inclination angle $\is$ is particularly
important to probe the spin-orbit alignment of transiting exoplanetary
systems in an independent and complementary manner to the projected
angle $\lambda$ from \RM measurement \citep{Ohta:2005aa}.  The
statistical distribution of $\is$ and $\lambda$ provides a
quantitative test for theories of origin and evolution of planetary
systems.

While the majority of transiting exoplanets are found around F, G, and
K type stars in their main-sequence phase, those are harder to measure
$\is$ compared to evolved solar-like stars (red-giants and subgiants).
This is mostly due to the relatively lower oscillation amplitude and
the sever mode blending of main-sequence solar-like stars. Therefore,
it is of fundamental importance to perform a systematic verification
of the reliability of $\is$ derived asteroseismology for those stars.

We generated 3000 simulated oscillation power spectra scaled from a
reference star KIC 12069424 (16 Cyg A) that span a wide range of the
height-to-background ratio, rotational splitting $\spl$, and
inclination angle $\is$, each for 1 year and 4 years observation
duration. Then we performed systematic mock simulations of
asteroseismic analysis, and examined the reliability of $\is$ derived
from asteroseismic analysis with a Bayesian-MCMC sampling method.

We find that the low signal-to-noise ratio of the power spectra
induces a systematic under-estimate (over-estimate) bias for stars
with high (low) inclinations. The combination of analytical
consideration and mock simulation results revealed three empirical
criteria on ($\is$, $\spl/\Gamma$) plane as a function of the power,
height-to-background ratio HBR, and the observation duration
$T_{\rm obs}$, which are required for a reliable estimate of $\is$.
The criteria indicate that reliable measurements are possible in
the range of $20^\circ \lesssim \is \lesssim 80^\circ$ for
stars with high HBR, high $\spl/\Gamma$, and/or longer $T_{\rm obs}$.

We also performed asteroseismic analysis of 94 main-sequence stars in
{\it Kepler} short cadence data using the same Bayesian-MCMC sampling
method; 33 and 61 are stars with and without known planetary
companions, respectively. We find that 9 stars with planet and 22
stars without planet satisfy the above criteria.

The stellar inclination and rotation, $\is$ and $\spl$, that we
derived asteroseismically for those {\it Kepler} stars are compared
with those derived photometrically and spectroscopically.  We find
that our asteroseismic $\vsini$ is in good agreement with the
{\it average} of two independent spectroscopic analysis by
\citet{Bruntt:2012aa} and \citet{Molenda-Zakowicz:2013aa}.  This
suggests that a careful modelling of macroturbulence is crucial in
estimating $\vsini$ from spectroscopic data, especially for slowly
rotating stars.

The rotation period $P_{\rm rot}$ derived from the photometric variability
of the stellar light curve shows reasonable, even if not good,
agreement with $\spl$.  The combined estimate of $\is$,
however, is very limited both observationally and statistically, and
does not show strong agreement with its asteroseismic estimate at this
point, indicating that further quantitative study is necessary.  The
statistical discussion and implications of our asteroseismic result
for the {\it Kepler} stars will be presented in a future study.


\section*{Acknowledgements}

We thank NASA {\it Kepler} team and KASOC teams for making their data
available to us.  We are grateful to Mikkel Lund and Tiago Campante,
who kindly shared our problems in simulation and gave us materials to
solve them, to Masataka Aizawa, Kento Masuda, Hajime Kawahara, Jerome
Ballot, Rafael Garcia, and Thierry Appourchaux for their valuable
comments on the direction of this work, and to Benoit Marchand and
Martin Nielsen for their technical support and help on data
preparation. We also thank Benoit Mosser, the referee of the paper,
for numerous constructive and valuable comments, which we hope
significantly improved the presentation and clarity of the earlier
manuscript.  The numerical computation was carried out on DALMA
cluster in New York University Abu Dhabi, and PC cluster at Center for
Computational Astrophysics, National Astronomical Observatory of
Japan.  S.K. is supported by JSPS (Japan Society for Promotion of
Science) Research Fellowships for Young Scientists (No. 16J03121).
Y.S. gratefully acknowledges the support from Grants-in Aid for
Scientific Research by JSPS No. 24340035. The present research is also
supported by JSPS Core-to-Core Program ``International Network of
Planetary Sciences''.



\bibliographystyle{mnras}





\bsp	
\label{lastpage}
\end{document}